\documentclass[%
 reprint,
 superscriptaddress,
 amsmath,amssymb,
 aps,
 pra,
 nofootinbib,
 showkeys,
 longbibliography
]{revtex4-2}

\usepackage{graphicx}
\usepackage{dcolumn}
\usepackage{enumitem}
\usepackage{bm}
\usepackage{physics}
\usepackage{tensor}
\usepackage{hyperref}
\hypersetup{
    colorlinks,
    linkcolor={blue},
    urlcolor={blue},
    citecolor={blue},
}
\usepackage{color}

\begin{document}

\preprint{APS/123-QED}

\title{Spectral damping without quasiparticle decay: The dynamic structure factor \texorpdfstring{\\}{} of two-dimensional quantum Heisenberg antiferromagnets}

\author{Matthew C. O'Brien}
\email{mco5@illinois.edu}
\affiliation{
School of Physics, The University of New South Wales, Sydney, New South Wales 2052, Australia
}
\affiliation{Department of Physics, University of Illinois at Urbana-Champaign, Urbana, Illinois 61801-3080, USA}
\author{Oleg P. Sushkov}%
\email{sushkov@unsw.edu.au}
\affiliation{
School of Physics, The University of New South Wales, Sydney, New South Wales 2052, Australia
}


\date{\today}

\begin{abstract}
Two-dimensional Heisenberg antiferromagnets play a central role in quantum magnetism, yet the nature of dynamic correlations in these systems at finite temperature has remained poorly understood for decades. We solve this long-standing problem by using a novel quantum-classical duality to calculate the dynamic structure factor analytically and, paradoxically, find a broad frequency spectrum despite the very long quasiparticle lifetime. The solution reveals new multi-scale physics whereby an external probe creates a classical radiation field containing infinitely-many quanta. Crucially, it is the multi-scale nature of this phenomenon which prevents a conventional renormalization group approach. We also challenge the common wisdom on static correlations and perform Monte Carlo simulations which demonstrate excellent agreement with our theory.
\end{abstract}

\maketitle

\section{Introduction \label{sec:intro}}

The dynamic structure factor encodes the fundamental physical processes involved in the response of a system to an external probe, and is the most common experimental observable in studies of magnetic systems, determined, for example, using inelastic neutron \cite{Banerjee2017}, or resonant x-ray scattering spectroscopy \cite{Halasz2016}. However, theoretical analyses of these processes which are both quantitatively accurate and physically insightful can be elusive. The two-dimensional quantum Heisenberg antiferromagnet (2DQHA) plays an important role in the field of quantum magnetism precisely because of the theoretical challenges it poses in addition to its descriptive power: First, the model describes the parent compounds of cuprate high temperature superconductors \cite{Kastner1998}. Second, while the model supports long-range order at zero temperature, order is destroyed at any finite temperature \cite{Mermin1966}. Because of the importance of thermal fluctuations, 2DQHAs manifest highly non-trivial classical and quantum long-range dynamics which are not fully understood \cite{zinn2002quantum,Sachdev2011}. The nature of quantum critical points to and from quantum spin liquid phases is also a problem of intense theoretical interest (see Ref. \cite{Savary2017} for a review). Somewhat surprisingly, the physics of thermal fluctuations in isotropic 2DQHAs is closely related to the zero temperature quantum Lifshitz phase transition between antiferromagnetically ordered states and a spin liquid phase in systems with long-range frustrated interactions (e.g., the $J_1$-$J_3$ model) \cite{OBrien2020a}. 

In their seminal work, Chakravarty, Halperin and Nelson used the $O(3)$-symmetric nonlinear $\sigma$ model (NLSM) to describe the long wavelength physics of 2DQHAs at low temperature and argued that the spin-spin correlations in the so-called ``renormalized classical'' regime are essentially classical in nature \cite{Chakravarty1989}. Crucially, their analysis relied on a quantum-classical mapping which integrates out all dynamics of the quantum model. Consequently, this approach allowed the authors to derive a scaling form for the static structure factor but not for the dynamic structure factor. Later studies of the dynamic structure factor raised surprising questions. First, a direct perturbative calculation of the magnon decay rate due to scattering from the thermal bath predicted the dynamic structure factor should have a very narrow linewidth \cite{Ty1990}. Similarly, a $1/N$ expansion of the $O(N)$ NLSM predicted a narrow quasi-Lorentzian frequency distribution \cite{Chubukov1994}. In contrast, classical time-dependent numerical simulations showed a broad dynamic structure factor \cite{Tyc1989}, and it has so far remained unclear how to rigorously reconcile this apparent contradiction.

We resolve the long-standing discrepancy in this paper with a novel analytical calculation of the dynamic spin structure factor of the isotropic $O(N\geq3)$ NLSM at finite temperature. In recent works \cite{OBrien2020,OBrien2020a}, we demonstrated that infrared-divergent fluctuations---either thermal or quantum---lead to the emergence of a new quantum-classical duality: when an external probe interacts with the system, it creates a classical field which contains an infinite number of quanta with finite total energy. This concept actually originates from particle physics where it was first developed by Bloch and Nordsieck to solve the problem of the radiation field of accelerating electrons \cite{Bloch1937}. Since the $O(2)$ NLSM is exactly solvable, we were able to rigorously show that despite the infinite quasiparticle lifetime, the dynamic structure factor at nonzero temperature is broad and non-Lorentzian \cite{OBrien2020}.

The $O(N\geq3)$ models are not exactly solvable, and hence, the diagrammatic expansion we derived in Ref. \cite{OBrien2020} is not applicable. However, we leverage the concept of the infrared catastrophe to develop a new analytical technique and use it to show for the first time that the dynamic spin structure factor of the $O(N)$ quantum NLSM at finite temperature is very broad and non-Lorentzian. Our analysis demonstrates that this broadening is not due to short-lived quasiparticles but instead is due to the radiation of multiple spin waves by the external probe. With this result, we also obtain the static structure factor by integrating over frequency and find similarities with the scaling form known in the literature \cite{Chakravarty1989,Chubukov1994,Sachdev2011}. However, our static structure factor has a different temperature dependence which originates from the underlying quantized nature of the highly-classical radiation field. Fortunately, unlike the dynamic factor, the static structure factor can be calculated numerically using path integral quantum Monte Carlo---referred to hereafter as Monte Carlo (MC). Therefore, to confirm our result for the static structure factor we also perform extensive MC simulations of the $O(3)$-symmetric NLSM and find excellent agreement.

The rest of this paper is structured as follows: In Sec. \ref{sec:formalism}, we introduce the nonlinear $\sigma$ model as the effective field theory for the 2D quantum Heisenberg antiferromagnet, and summarize some of its most important features, including the renormalization group running coupling constants. In Sec. \ref{sec:DSF}, we present the main result of this paper: a calculation of the dynamic structure factor of the $O(N)$ NLSM which accounts for the non-trivial multi-scale physics. In Sec. \ref{sec:equaltime}, we use our new expression for the dynamic structure factor to study the equal-time correlations of the NLSM, and present the results of MC simulations which support our predictions. Section \ref{sec:summary} presents our conclusions.


\section{Formalism \label{sec:formalism}}

\begin{figure}[t]
    \centering
    \includegraphics[scale=0.6]{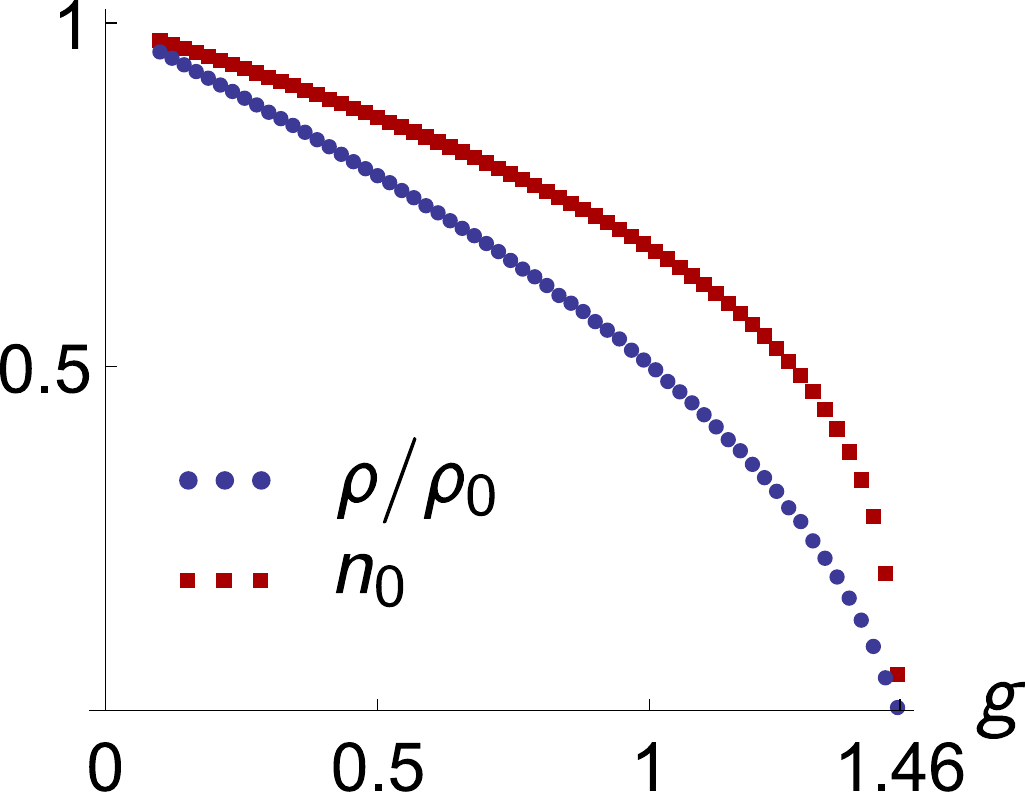}
    \caption{Zero temperature spin stiffness $\rho$ and average staggered magnetization $n_0$ of the $O(3)$ NLSM as a function of $g=\hbar c/\rho_0 b$ measured using MC on a $64^3$ lattice. Both $\rho$ and $n_0$ vanish at the critical point $g = g_c \approx 1.46$.}
    \label{fig:rho_mag}
\end{figure}

The long-range dynamics of 2DQHAs at low temperature can be described by the $O(3)$ NLSM with Lagrangian $\mathcal{L} = (\rho_0/2) [c^{-2} (\partial_t\Vec{n}_0)^2 - (\grad \Vec{n}_0)^2]$, $\Vec{n}_0^2 = 1$, where $\rho_0$ and $\Vec{n}_0$ are the spin stiffness and staggered magnetization order parameter, respectively, defined at the ultraviolet scale $\Lambda_0 \sim \pi/b$, and $b$ is the lattice spacing \cite{Chakravarty1989,Ioffe1988}. Quantum fluctuations are ultraviolet-divergent as a power of the momentum scale, and at a scale $\Lambda_1 \ll \Lambda_0$ corresponding to several lattice spacings, reduce the order parameter down to $n_0 = \lvert \langle \Vec{n}_0 \rangle \rvert < 1$. To describe physics at the scale $\Lambda_1$, the quantum fluctuations can be integrated out, leading to the low energy Lagrangian
\begin{equation}
    \mathcal{L} = \frac{1}{2} \rho \left[\frac{1}{c^2}(\partial_t \Vec{n})^2 - (\grad \Vec{n})^2\right] ,\qquad \Vec{n}^2 = 1, \label{eq:lagrangian}
\end{equation}
where $\rho$ and $\Vec{n}$ are the spin stiffness and order parameter normalized at $\Lambda_1$. Both $\rho$ and $n_0$ as a function of the dimensionless coupling constant $g = \hbar c/\rho_0 b$ can be calculated numerically using MC, and those of the $O(3)$ NLSM are shown in Fig. \ref{fig:rho_mag}. In this paper, we address the regime $g < g_c\approx 1.46$ which describes 2DQHAs \cite{Chakravarty1989}.

For the sake of generality, we consider from hereon the $O(N)$-symmetric model in terms of $\Vec{n} = (n^1, n^2, \dots, n^N)$, with $N\geq 3$, and use units of $\hbar=c=b=1$. At zero temperature, the ground state of the model has long-range collinear antiferromagnetic order---$\Vec{n}=\text{const}$---which spontaneously breaks the $O(N)$ symmetry. However, the Hohenberg-Mermin-Wagner theorem guarantees the destruction of long-range order at any finite temperature \cite{Mermin1966,Hohenberg1967}. Despite this, at sufficiently low temperatures $T \ll \rho$, the system remains ordered on scales up to the exponentially-large correlation length $\xi \propto \exp[2\pi \rho/(N-2)T]$ \cite{Chakravarty1989,Chubukov1994,Sachdev2011,Hasenfratz1990a}. This separation of scales implies a notion of quasi-long-range order and allows for a perturbative treatment of the NLSM on momentum scales $\Lambda$ satisfying $\xi^{-1} \ll \Lambda \leq \Lambda_1$. For momentum scales on the order of temperature to fall within this range, it suffices for $T\ll\rho$. The effects of fluctuations on scales $\Lambda \sim T$ can be determined within the leading order of perturbation theory. However, there are two types of contributions governing the physics of fluctuations on scales $\Lambda < T$: (i) Renormalization group (RG) ``running'' of physical parameters due to interactions occurring at the same scale. These are the conventional interactions which are well-understood in field theory. (ii) Contributions originating from multi-scale interactions to which conventional RG techniques are blind; we refer to these as the ``beyond RG'' contributions. The general principle underlying this paper is the novel technique we have developed to unify these two different contributions. 

The RG contributions are well-understood \cite{Chakravarty1989,Sachdev2011,zinn2002quantum}, so we summarize only the general principles here. The unit vector constraint of Eq. \eqref{eq:lagrangian} generates interactions between the components of $\Vec{n}$, leading to renormalization of the spin stiffness---$\rho \rightarrow \rho_\Lambda$---and fields---$\Vec{n} \rightarrow \Vec{n}_\Lambda = Z_\Lambda^{1/2} \Vec{n}$, where $Z$ is the quasiparticle residue. To one-loop accuracy at the momentum scale $\Lambda < T < \Lambda_1$,
\begin{subequations} \label{eq:RG_equations}
\begin{align}
    \rho_\Lambda &= \rho - \frac{(N-2)T}{2\pi} \log \frac{T}{\Lambda}, \label{eq:RG_rho} \\
    Z_\Lambda &= \frac{1}{n_0^2}\left(\frac{\rho}{\rho_\Lambda}\right)^{\textstyle \frac{N-1}{N-2}}.
\end{align}
\end{subequations}
The ultraviolet cutoff for the fluctuations in \eqref{eq:RG_rho} is the temperature $T$ rather than $\Lambda_1$ due to the bosonic statistics of the quasiparticles; this is an important quantum correction to classical thermodynamics \cite{Nelson1977}. In Appendix \ref{app:formalism}, we give a more detailed derivation of \eqref{eq:RG_equations} and show that higher-loop contributions are negligible when $T\ll\rho$.

\section{Dynamic Structure Factor \label{sec:DSF}}

The dynamic structure factor (DSF) is the Fourier transform of the order parameter correlation function\footnote{Note that in a previous work \cite{OBrien2020}, we defined the structure factor as the Fourier transform of $\langle \Vec{n}(\mathbf{r},t)\cdot\Vec{n}(0)\rangle$, the total response from all polarizations. The present definition differs by a factor of $1/N$.}
\begin{equation}
    S(\mathbf{k},\omega) \delta_{ij} = \int d t\, d^2 \mathbf{r}\, \langle n^i(\mathbf{r},t) n^j(0) \rangle\, e^{i(\omega t - \mathbf{k}\cdot \mathbf{r})}, \label{eq:skw}
\end{equation}
where the average is taken over the thermal ensemble, and is independent of the polarization indices $i$, $j$ due to the absence of long-range order at finite temperature. Expanding \eqref{eq:skw} in a spectral representation in the basis of excited quasiparticle Fock states $\ket{\alpha}$ and $\ket{\beta}$ yields \cite{Lifshitz1995},
\begin{align}
    S(\mathbf{k},\omega) &= \sum_{\alpha,\beta} \frac{e^{-\omega_\beta/T}}{\mathcal{Z}} \lvert \bra{\alpha} n^{i}(0) \ket{\beta} \rvert^2 \nonumber\\
    & \times (2\pi)^3 \delta(\omega - \omega_\alpha + \omega_\beta)\delta^{(2)}(\mathbf{k} - \mathbf{k}_\alpha + \mathbf{k}_\beta), \label{eq:skw_spectral}
\end{align}
where $\mathcal{Z}$ is the quantum partition function, and $\omega_\alpha$ and $\mathbf{k}_\alpha$ are the energy and momentum of the state $\ket{\alpha}$. In this paper we always work with $\omega > 0$, since \eqref{eq:skw_spectral} implies that $S(\mathbf{k},-\omega) = e^{-\omega/T}S(\mathbf{k},\omega)$.

\subsection{RG contributions}

\begin{figure}[t]
    \centering
    \includegraphics[scale=0.6]{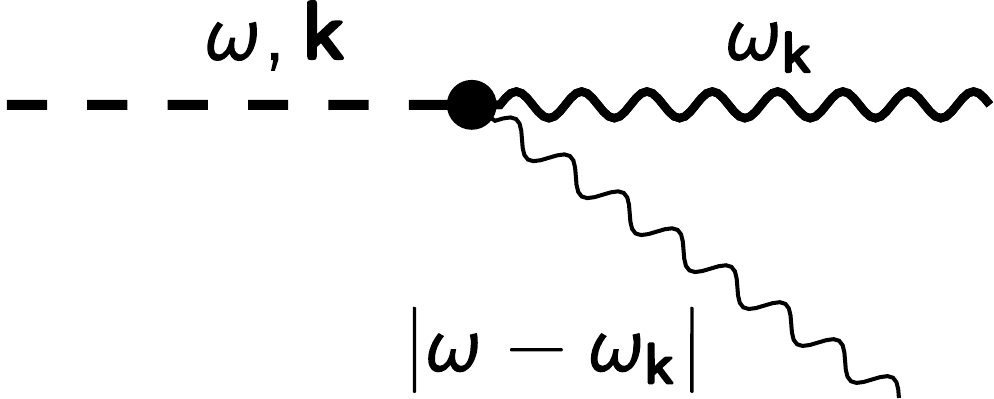}
    \caption{The dominant contribution to the dynamic structure factor is the emission by the probe (dashed line) of a quasiparticle with energy $\omega_{\mathbf{k}}$ (thick wavy line) and a second ``soft'' particle with energy $\lvert\omega - \omega_{\mathbf{k}} \rvert \ll \omega_{\mathbf{k}}$ (thin wavy line).}
    \label{fig:emit}
\end{figure}

First, we account for RG contributions to the DSF by renormalizing the fields in \eqref{eq:skw_spectral} at the scale of the incoming momentum $\mathbf{k}$, so that $n^i_{\mathbf{k}} = n^i/Z_{\mathbf{k}}^{1/2}$; since $\xi^{-1} \ll k < T$, local order exists at this scale. The absence of long-range order means that the single magnon intermediate state does not contribute to \eqref{eq:skw_spectral} \cite{OBrien2020}. Therefore, we eliminate the $\Vec{n}^2 = 1$ constraint by writing the order parameter field as $\Vec{n}_{\mathbf{k}} = (\Vec{\pi}_{\mathbf{k}}, \sqrt{1-\smash[b]{\Vec{\pi}_{\mathbf{k}}^2}})$, where $\Vec{\pi} = (\pi^1,\dots,\pi^{N-1})$ are small transverse fluctuations, and use the component of $\Vec{n}$ in the direction of local order---$n^N_{\mathbf{k}} = \sqrt{1-\smash[b]{\Vec{\pi}_{\mathbf{k}}^2}}$---to compute the DSF. However, this approach assigns all dynamics to the directions transverse to the local moment, and hence, does not respect the $O(N)$ symmetry which must remain unbroken at finite temperature. To restore symmetry, we rotationally average the DSF over all $N$ polarizations by multiplying \eqref{eq:skw_spectral} by $(N-1)/N$ \cite{Chubukov1994,Sachdev2011}. Therefore, suppressing Boltzmann factors and $\delta$ functions for notational clarity, the DSF is
\begin{equation}
    S(\mathbf{k},\omega) = \left(\frac{N-1}{N}\right) \frac{1}{Z_{\mathbf{k}}} \sum_{\alpha,\beta} \lvert \bra{\alpha} \sqrt{1 - \smash[b]{\Vec{\pi}_{\mathbf{k}}^2}} \ket{\beta} \rvert^2. \label{eq:skw_zcomp}
\end{equation}
We have primarily chosen the longitudinal component of $\Vec{n}$ to simplify the calculations which follow. In Appendix \ref{app:transverse_comparison}, we describe the mechanism which is responsible for restoring rotational symmetry in this representation of the order parameter. The leading contribution is then obtained by expanding $\sqrt{1 - \smash[b]{\Vec{\pi}_{\mathbf{k}}^2}} \simeq 1 - \smash[b]{\Vec{\pi}_{\mathbf{k}}^2}/2$. Naively, the first term appears to yield a $\mathbf{k} = 0$ Bragg peak. However, we emphasize that order only \textit{locally} exists on scales $0 < \xi^{-1} < k$, so it is incorrect to use \eqref{eq:skw_zcomp} for momenta $k < \xi^{-1}$. Therefore, we take $1 - \smash[b]{\Vec{\pi}_{\mathbf{k}}^2}/2 \rightarrow - \smash[b]{\Vec{\pi}_{\mathbf{k}}^2}/2$ and evaluate the leading contribution by using Fermi's golden rule to find the probability of two magnon radiation. More precisely, if $\omega > \omega_{\mathbf{k}}$ the external probe excites two quasiparticles (see Fig. \ref{fig:emit}), and if $\omega < \omega_{\mathbf{k}}$ one quasiparticle is emitted and a second is absorbed. When $\lvert \omega - \omega_{\mathbf{k}} \rvert \doteq \lvert\Delta\rvert \ll \omega_{\mathbf{k}}$ both processes have the same contribution to the sum over initial and final states 
\begin{align}
    \Tilde{I}_2(\mathbf{k},\omega) &\simeq \frac{N-1}{2!} \int \frac{d^2\mathbf{k}_1}{(2\pi)^2} \frac{d^2\mathbf{k}_2}{(2\pi)^2} \frac{T}{2\omega_1^2 \rho_{\mathbf{k}}} \frac{T}{2\omega_2^2 \rho_{\mathbf{k}}} \nonumber \\
    & \times (2 \pi)^3 \delta(\omega - \omega_1 - \omega_2) \delta^{(2)}(\mathbf{k} - \mathbf{k}_1 - \mathbf{k}_2) \nonumber \\
    &= \frac{(N-1)T^2}{4\rho_{\mathbf{k}}^2\omega_{\mathbf{k}}^2 \lvert \omega - \omega_{\mathbf{k}} \rvert}, \label{eq:2partRG}
\end{align}
where the spin stiffness $\rho_{\mathbf{k}}$ is evaluated at $\mathbf{k}$ due to our choice of renormalization scale.

However, by examining the structure of the phase space integral above, we find that one emitted quasiparticle will have energy $\sim \omega_{\mathbf{k}}$ and the other will have energy $\sim \lvert\Delta\rvert \ll \omega_{\mathbf{k}}$. Hence, the two magnon intermediate state is an inherently multi-scale process and contributions at the ``soft'' scale $\Lambda \sim \lvert\Delta\rvert$ are not properly accounted for; conventional RG is not sufficient to describe the process accurately. 


\subsection{Beyond RG contributions}

We now account for the multi-scale nature of the two magnon intermediate state. First, observe that we can perform a \textit{post hoc} simplification of the phase space integral leading to \eqref{eq:2partRG} using our knowledge of the momentum distribution, and find that it factorizes as
\begin{align}
    \tilde{I}_2(\mathbf{k},\omega) &\simeq (N-1)  \left(\int \frac{d^2 \mathbf{k}_1}{(2\pi)^2} \frac{T}{2 \omega_1^2 \rho_{\mathbf{k}}} (2\pi)^2\delta^{(2)}(\mathbf{k} - \mathbf{k}_1)\right) \nonumber \\
    & \times \left(\int \frac{d^2 \mathbf{k}_2}{(2\pi)^2} \frac{T}{2 \omega_2^2 \rho_{\mathbf{k}}} (2\pi)\delta(\Delta - \omega_2) \right), \label{eq:2part_factorized}
\end{align}
into high and low energy processes. This implies that to leading order, the RG corrections to the properties of the emitted quasiparticles will also factorize. Importantly, we can then account for the running of the parameters---the quasiparticle residue and the spin stiffness---of the two particles independently at their respective momentum scales. This leads to an \textit{explicit} dependence of the spin stiffness on the magnon momentum $\rho \rightarrow  \rho(\mathbf{k}_j)$---which we denote in this manner to emphasize that the momentum argument is now a variable---while the running of the quasiparticle residue from the normalization point $\mathbf{k}$ down to the magnon momentum gives an additional factor of $Z_{\mathbf{k}}/Z(\mathbf{k}_j)$ for each particle. Note that the additional factors of $Z_{\mathbf{k}}$ in the numerator come from the fact that the spectral expansion of the DSF \eqref{eq:skw_zcomp} is already normalized at the momentum transfer $\mathbf{k}$ from the external probe, so there is no need to account a second time for the running from the ultraviolet down to this scale. Therefore, the beyond RG version of \eqref{eq:2partRG}---denoted with no tilde---is straightforward to evaluate to logarithmic accuracy
\begin{align}
     I_2(\mathbf{k},\omega) &\simeq \frac{N-1}{2!} \int \frac{d^2 \mathbf{k}_1}{(2\pi)^2} \frac{d^2 \mathbf{k}_2}{(2\pi)^2} \frac{T Z_{\mathbf{k}}}{2 \omega_1^2 \rho(\mathbf{k}_1) Z(\mathbf{k}_1)} \nonumber \\
    &\mkern-58mu \times\frac{T Z_{\mathbf{k}}}{2 \omega_2^2 \rho(\mathbf{k}_2) Z(\mathbf{k}_2)} (2 \pi)^3 \delta(\omega - \omega_1 - \omega_2) \delta^{(2)}(\mathbf{k} - \mathbf{k}_1 - \mathbf{k}_2) \nonumber \\
    &\simeq \frac{Z_{\mathbf{k}}}{Z_\Delta} \frac{(N-1)T^2}{4\rho_{\mathbf{k}} \rho_\Delta \omega_{\mathbf{k}}^2 \lvert \omega - \omega_{\mathbf{k}} \rvert} . \label{eq:2part_corrected}
\end{align}
We note that there are two-loop corrections to the source vertex which do not factorize. However, this will be only a higher-order effect, and so does not influence our current discussion of the physics at leading order.


We understand from our exact solution of the $O(2)$ NLSM that the physics of the soft scale $\Lambda \sim \lvert\Delta\rvert$ is characterized by an interplay between thermal fluctuations and the radiation of additional (more than two) arbitrarily low energy quasiparticles, the ``probabilities'' of both of which are logarithmically infrared-divergent; this divergence implies that no finite number of quasiparticles can be excited by the probe \cite{OBrien2020}. Therefore, the ``second quasiparticle'' with energy $\lvert\Delta\rvert$ emitted/absorbed by the probe is actually accompanied by a classical radiation field containing infinitely-many quanta. However, mathematically, this radiation field is indistinguishable from the long-wavelength static thermal fluctuations, and serves only to set an infrared cut-off at the smallest physically relevant momentum scale $\Lambda_{\mathrm{min}} = \lvert\Delta\rvert$. This is a peculiarity of the Bloch-Nordsieck physics which has been long-known in particle physics: Accelerating matter always emits an infinite number of gauge quanta, whether it be electrons emitting photons \cite{Bloch1937}, or generic massive particles emitting gravitons \cite{Weinberg1965}. Therefore, having already accounted for the multi-scale nature of two magnon radiation down to $\lvert\Delta\rvert$, we have implicitly resummed all the leading large logarithmic contributions from higher-order processes involving more than two magnons in the intermediate state.

Therefore, we find that the DSF of the $O(N)$ NLSM in the regime $\xi^{-1} \ll \lvert\Delta\rvert \ll \omega_{\mathbf{k}} \ll T$ is
\begin{align}
    S(\mathbf{k},\omega) &= \left(\frac{N-1}{N}\right) \frac{1}{Z_{\mathbf{k}}} I_2(\mathbf{k},\omega) \nonumber \\
    &= \frac{(N-1)^2}{N} \frac{1}{Z_\Delta} \frac{T^2}{4\rho_{\mathbf{k}}\rho_{\Delta} \omega_{\mathbf{k}}^2 \lvert \omega - \omega_{\mathbf{k}} \rvert}, \label{eq:skw_compact}
\end{align}
so that in terms of the one-loop expressions for $Z_{\Delta}$, $\rho_{\mathbf{k}}$ and $\rho_\Delta$ given by \eqref{eq:RG_equations}, the full form of \eqref{eq:skw_compact} is
\begin{widetext}
\begin{equation}
    S(\mathbf{k},\omega) = \frac{(N-1)^2}{N} \left[ 1 - \frac{(N-2)T}{2\pi\rho}\log\frac{T}{\omega_{\mathbf{k}}} \right]^{-1} \left[1 - \frac{(N-2)T}{2\pi\rho}\log\frac{T}{\lvert\omega - \omega_{\mathbf{k}}\rvert}\right]^{\textstyle \frac{1}{N-2}} \frac{T^2 n_0^2}{4 \rho^2 \omega_{\mathbf{k}}^2 \lvert \omega - \omega_{\mathbf{k}} \rvert} , \label{eq:skw_full}
\end{equation}
It is common to express the structure factor in terms of appropriate length/time scales. In the present case, the only length scale is
\begin{equation}
    \lambda = \frac{1}{T} \exp\left[ \frac{2\pi\rho}{(N-2)T} \right], \label{eq:lambda_scale}
\end{equation}
in which case we can write
\begin{equation}
    S(\mathbf{k},\omega) = \frac{(N-1)^2}{N} \frac{2 \pi \rho}{(N-2)T \log(\lambda \omega_{\mathbf{k}})} \left[\frac{(N-2)T}{2\pi\rho}\log(\lambda\lvert\omega - \omega_{\mathbf{k}}\rvert)\right]^{\textstyle \frac{1}{N-2}} \frac{T^2 n_0^2}{4 \rho^2 \omega_{\mathbf{k}}^2 \lvert \omega - \omega_{\mathbf{k}} \rvert} . \label{eq:skw_lambda}
\end{equation}\\
\end{widetext}

Of course, this result assumes $\lambda\lvert\omega - \omega_{\mathbf{k}}\rvert \gg 1$, and hence, represents a very broad frequency distribution decaying more slowly than $1/\lvert\Delta\rvert$. The limit $N\rightarrow2$ reproduces the exact solution obtained in Ref. \cite{OBrien2020} using a direct summation of diagrams; this is clearly one important source of validation of our present approach. While we used a very different technique in Ref. \cite{OBrien2020}, the hierarchy of multi-particle contributions was compatible with the present approach. Since the case $N > 2$ is not exactly solvable, in this work we used the running of parameters to correctly account for the beyond RG contributions.

\subsection{Lifetime damping}

We have so far neglected the lifetime of quasiparticles in the $O(N\geq3)$ NLSM. The dominant decay process for an on-shell magnon with energy $\omega_{\mathbf{k}} \ll T$ is $2\rightarrow2$ Raman scattering from a particle in the thermal bath, which leads to the well known inverse lifetime\footnote{Here $\Gamma_{\mathbf{k}}$ is the full width at half maximum of the quasiparticle Green's function. Some works define it as the \textit{half} width, leading to a spurious factor of 2 difference.} \cite{Chubukov1994,Ty1990},
\begin{figure}[!t]
    \centering
    \includegraphics[scale=1.2]{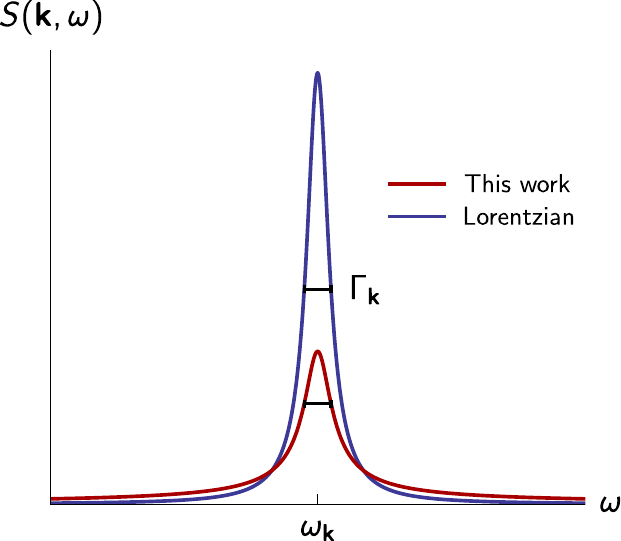}
    \caption{The dynamic structure factor \eqref{eq:skw_lifetime} accounting for lifetime broadening with $\omega_{\mathbf{k}} = T/2 = \rho/4$. The Lorentzian FWHM $\Gamma_{\mathbf{k}}/\omega_{\mathbf{k}} \approx 0.015$ is given by \eqref{eq:linewidth}. Both curves have the same $\omega$ integrate spectral weight.}
    \label{fig:spectrum}
\end{figure}
\begin{equation}
    \frac{\Gamma_{\mathbf{k}}}{\omega_{\mathbf{k}}} \simeq \frac{(N-2)T^2}{4\pi\rho_{\mathbf{k}}^2} \log\frac{T}{\omega_{\mathbf{k}}} . \label{eq:linewidth}
\end{equation}
Importantly, in our regime of interest ($\xi^{-1} \ll \omega_{\mathbf{k}} \ll T \ll \rho$) $\Gamma_{\mathbf{k}}/\omega_{\mathbf{k}}$ is an $\mathcal{O}(T^2/\rho^2)$ small quantity. It is then clear from the analysis in the previous subsection that when $\Gamma_{\mathbf{k}} < \lvert\Delta\rvert$, radiative broadening of the DSF due to multiple emissions/absorptions dominates over $1/\lvert\Delta\rvert^2$ Lorentzian lifetime broadening. As a side note, since $\Gamma_{\mathbf{k}} \propto N$, radiative broadening may be hidden in a $1/N$ expansion around $N = \infty$. However, for $N$ not much larger than $\mathcal{O}(\rho/T)$, the region $\lvert\Delta\rvert < \Gamma_{\mathbf{k}}$ remains very narrow. Regardless, lifetime broadening cannot be neglected near resonance. 

The finite lifetime has the effect of ``broadening'' the energy conserving $\delta$ function in \eqref{eq:2part_corrected}
\begin{align}
    (2 \pi) \delta(\omega - \omega_1 - \omega_2) &\rightarrow \frac{\Gamma}{(\omega - \omega_1 - \omega_2)^2 + \Gamma^2/4} \nonumber \\
    &\simeq \frac{\Gamma_{\mathbf{k}}}{(\Delta - \omega_2)^2 + \Gamma_{\mathbf{k}}^2/4},
\end{align}
since $\Gamma_{\Delta} \ll \Gamma_{\mathbf{k}}$. However, there is an important subtlety in accounting for contributions from different pieces of phase space. Without lifetime broadening, the following cases are possible:
\begin{enumerate}[label=(\roman*),labelindent=\parindent]
    \item $\omega > \omega_{\mathbf{k}} > 0$: two particles are emitted.
    \item $\omega_{\mathbf{k}} > \omega > 0$: one particle is emitted with energy $\omega_{\mathbf{k}}$ and one is absorbed with energy $\lvert\Delta\rvert$.
    \item $0 > \omega > -\omega_{\mathbf{k}}$: one particle is absorbed with energy $\omega_{\mathbf{k}}$ and one is emitted with energy $\lvert\Delta\rvert$.
    \item $0 > -\omega_{\mathbf{k}} > \omega$: two particles are absorbed.
\end{enumerate}
In principle, with account of lifetime broadening, any of these processes can occur for any values of energy and momentum transfer from the source. However, since we work in the regime $\lvert\Delta\rvert \ll \omega_{\mathbf{k}}$, we can safely assume no mixing between the positive and negative frequency branches of the spectrum. However, for $\omega > 0$, we must allow for mixing between processes (i) and (ii). Therefore, we generalize the integral \eqref{eq:2part_corrected} to give us the full form of the DSF
\begin{widetext}
\begin{equation}
    S(\mathbf{k},\omega) \simeq \frac{(N-1)^2}{N} \frac{T^2 n_0^2}{4 \rho_{\mathbf{k}} \omega_{\mathbf{k}}^2 } \int_{1/\lambda}^{\omega_{\mathbf{k}}} \frac{d^2 \mathbf{q}}{(2\pi)^2} \frac{1}{\omega_{\mathbf{q}} \rho(\mathbf{q}) Z(\mathbf{q})} \left[ \frac{\Gamma_{\mathbf{k}}}{(\Delta - \omega_{\mathbf{q}})^2 + \Gamma_{\mathbf{k}}^2/4} + \frac{\Gamma_{\mathbf{k}}}{(\Delta + \omega_{\mathbf{q}})^2 + \Gamma_{\mathbf{k}}^2/4} \right], \label{eq:skw_lifetime}
\end{equation}
\end{widetext}
which is plotted in Fig. \ref{fig:spectrum}, where it is compared to a Lorentzian lineshape with the same $\omega$ integrated spectral weight. Clearly, the resonant response of the DSF is greatly suppressed compared to the Lorentzian, with significant spectral weight shifted to the tails of the frequency distribution. Note that we must retain an infrared momentum cutoff for this expression. Given that we have already established that the characteristic momentum scale of the DSF in this regime is $\lambda^{-1}$, we use this as the cutoff. For process (i), the two emitted particles are indistinguishable bosons, but we distinguish between them, so we must impose the ultraviolet cutoff $\omega_{\mathbf{k}}$ to avoid double counting states. For process (ii), the dominant contribution comes from the absorption of particles with energy $< \omega_{\mathbf{k}}$. Finally, we note that in the limit $\lvert\Delta\rvert \gg \Gamma_{\mathbf{k}}$, \eqref{eq:skw_lifetime} reduces to the expression with no account of the lifetime \eqref{eq:skw_full}, as we anticipated in our discussion above.


\section{Equal-time correlations \label{sec:equaltime}}

The static structure factor can be calculated directly from the DSF by integrating over frequency. Since the resonant peak of the spectrum has a non-trivial spectral weight---as indicated by the non-integrable singularity at $\omega = \omega_{\mathbf{k}}$ in the zero lifetime expression \eqref{eq:skw_lambda}---we must use the full integral form \eqref{eq:skw_lifetime}. Taking note of the fact that $S(\mathbf{k}, -\omega) = e^{-\omega/T} S(\mathbf{k},\omega) \simeq S(\mathbf{k},\omega)$ when $\omega \ll T$, we find that
\begin{align}
    S(\mathbf{k}) &\simeq 2\int_0^\infty \frac{d \omega}{2 \pi} S(\mathbf{k},\omega) \nonumber \\
    &\simeq \left(\frac{N-1}{N} \right) \frac{T n_0^2}{\rho k^2} \left[\frac{(N-2)T}{2\pi\rho}\log(\lambda k)\right]^{\textstyle \frac{1}{N-2}} . \label{eq:sk}
\end{align}
We can also verify the total sum rule: Since the DSF we derived was valid for $\omega \ll T$, we should integrate \eqref{eq:sk} up to $T$:
\begin{equation}
    \int \frac{d^2 \mathbf{k}}{(2 \pi)^2} S(\mathbf{k}) \simeq \int_{1/\lambda}^T \frac{d^2 \mathbf{k}}{(2 \pi)^2} S(\mathbf{k}) = \frac{n_0^2}{N}.
\end{equation}
Therefore, summing up over the $N$ polarizations, we recover the correct normalization of the order parameter. We note that this sum rule is not satisfied if all parameters are normalized at the same scale---either $\omega_{\mathbf{k}}$ or $\lvert\Delta\rvert$. This observation further validates our approach to including multi-scale physics.\\

It then follows that the equal-time order parameter correlation function, which is $N$ times the Fourier transform of \eqref{eq:sk}, will be
\begin{align}
    \langle \Vec{n}(\mathbf{r})\cdot\Vec{n}(0)\rangle &= N \int \frac{d^2\mathbf{k}}{(2\pi)^2} S(\mathbf{k}) e^{-i \mathbf{k}\cdot\mathbf{r}} \nonumber \\
    &\simeq n_0^2 \left[ \frac{(N-2)T}{2\pi\rho} \log\left(\frac{\lambda}{r}\right) \right]^{\textstyle \frac{N-1}{N-2}} . \label{eq:equal_time_corr}
\end{align}

The static structure factor \eqref{eq:sk} has the same functional $k$ dependence as the well-known scaling form \cite{Chakravarty1989,Chubukov1994,Sachdev2011}. However, \eqref{eq:sk} contains $\log(\lambda k)$ instead of $\log(\xi k)$ in those references, where $\xi$ is the correlation length
\begin{equation}
    \xi = \frac{\xi_0}{T} \left[\frac{(N-2)T}{2\pi\rho}\right]^{\textstyle \frac{1}{N-2}} \exp\left[ \frac{2\pi\rho}{(N-2)T} \right], \label{eq:xi_length}
\end{equation}
$\xi_0 = (e/8)^{1/(N-2)}\Gamma[1 + 1/(N-2)]$, and $\Gamma(x)$ is the gamma function \cite{Chubukov1994,Sachdev2011,Hasenfratz1990a}. The replacement $\xi \rightarrow \lambda$ leads to a particularly drastic difference for the case $N=3$, where the pre-exponential factor of the correlation length is temperature-independent. 

\begin{figure}[!t]
    \centering
    \includegraphics[scale=1.2]{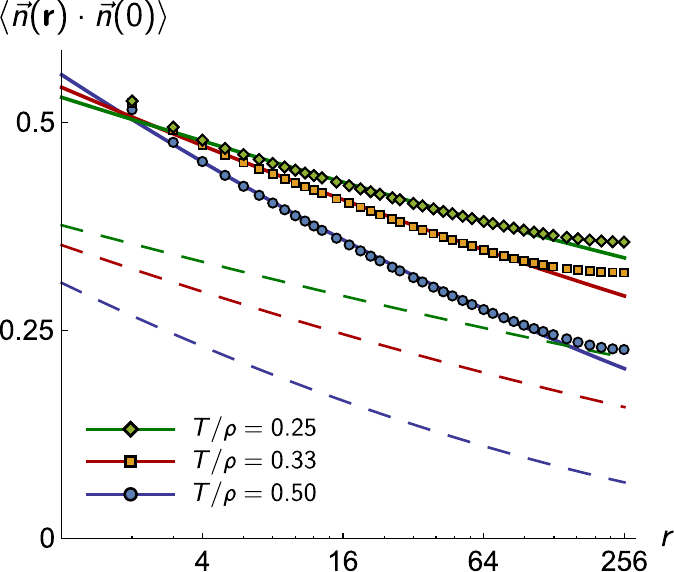}
    \caption{Order parameter equal-time correlations at fixed $g=1$ ($\rho \approx 0.504$ and $n_0 \approx 0.673$). Symbols are Monte Carlo data, solid lines are theory \eqref{eq:equal_time_corr} for $N=3$, and dashed lines are theory replacing $\lambda \rightarrow\xi$.}
    \label{fig:correlations}
\end{figure}

To confirm our results \eqref{eq:sk} and \eqref{eq:equal_time_corr}, we performed MC simulations of the $O(3)$ NLSM and measured the equal-time order parameter correlation function. The zero temperature spin stiffness $\rho$ and staggered magnetization $n_0$ presented in Fig. \ref{fig:rho_mag} have been calculated on a $64^3$ size lattice. To measure the correlation function we used lattices with $L_x = L_y = 512$ and $L_\beta$ = 4, 6, 8 imaginary time slices which correspond to different temperatures $T = g\rho_0/L_\beta$. In Fig. \ref{fig:correlations} we present the MC correlation function for dimensionless coupling $g = 1$, corresponding to $\rho/\rho_0 \approx 0.504$ and $n_0 \approx 0.673$. The solid lines show the theoretical prediction \eqref{eq:equal_time_corr} for $N=3$; note that the theory has \textit{no adjustable fitting parameters}. At $r \lesssim 2$---in units of the lattice spacing---deviations from theory are due to the dominance of ultraviolet quantum fluctuations on short length scales, and at $r \gtrsim 128$, finite-size effects from the periodic boundary conditions become important. Since the parameters used lie within the domain of validity of the theory, we find excellent agreement between our theoretical predictions and the MC simulation data. Further data and analysis can be found in Appendix \ref{app:MC}. The dashed lines show the correlation function \eqref{eq:equal_time_corr} with $\lambda$ replaced by $\xi$ and disagree very clearly with the MC simulations. 

To avoid misunderstanding we note the following: (i) The correlation length is defined in terms of the exponential decay of correlations on large length scales $\langle \Vec{n}(\mathbf{r})\cdot\Vec{n}(0)\rangle \sim e^{-r/\xi}$ when $r \gg \xi$. (ii) In this work, we are operating in the opposite limit $r \ll \xi$. We are not claiming that the well known expression \eqref{eq:xi_length} for the correlation length is incorrect. However, we claim that correlations on shorter scales are characterized by the parameter $\lambda$, and not $\xi$.

\section{Summary \label{sec:summary}}

We have calculated for the first time the finite temperature dynamic structure factor of the 2D $O(N)$ quantum nonlinear $\sigma$ model in the regime describing a Heisenberg antiferromagnet. The dynamic structure factor displays a very broad frequency distribution which decays more slowly than the first power of the detuning from resonance. Since the quasiparticle lifetime remains very long, it is irrelevant to the broad tails of the spectrum. Instead, the broadening is driven by the emission and absorption of multiple soft excitations by the probe. To perform this calculation, we developed a new analytical technique which accounts for both conventional single-scale renormalization group contributions and ``beyond RG'' effects from multi-scale physics. We expect this method to be broadly applicable to studying the dynamics of a wide range of finite temperature interacting quantum field theories.

Using our new result for the dynamic structure factor, we also calculated the static structure factor and found agreement of the functional momentum dependence with the previously known result. However, we predicted a significant modification of the characteristic length scale of correlations in the so-called scaling regime. This result implies an important correction to the temperature dependence of static correlations from the bosonic statistics of the quasiparticles. To confirm this prediction, we performed extensive path integral quantum Monte Carlo simulations and demonstrated perfect agreement between the numerical data and our analytical formula. To the best of our knowledge, this is also the first Monte Carlo study of correlations in the scaling regime.

\begin{acknowledgments}
We thank Jaan Oitmaa for consultations on Monte Carlo simulations, and Andrey Katanin, Michael Schmidt, and G\"otz Uhrig for important discussions. This research includes computations using the computational cluster Katana supported by Research Technology Services at UNSW Sydney. We have also received support from the Australian Research Council Centre of Excellence in Future Low Energy Electronics Technologies (CE170100039). 
\end{acknowledgments}

\appendix

\section{Formalism \label{app:formalism}}

\subsection{Finite temperature one-loop renormalization of the \texorpdfstring{($2+1$)}{(2+1)}-dimensional NLSM}

In Section \ref{sec:formalism}, we discussed how quantum and thermal fluctuations are taken into account via renormalization. Here we take a pedagogical approach to showing how the RG equations \eqref{eq:RG_equations} can be derived.

The $O(N)$ nonlinear $\sigma$ model (NLSM) in $2+1$ dimensions, normalized at the scale $\Lambda_0 \sim \pi/b$ where $b$ is the lattice spacing, is given by the Lagrangian
\begin{equation}
    \mathcal{L} = \frac{1}{2} \rho_0 (\partial_\mu \Vec{n}_0)^2,\qquad\qquad \Vec{n}_0^2 = 1, \label{eq:nlsm}
\end{equation}
where $\partial_\mu = (c^{-1}\partial_t,\partial_x,\partial_y)$ and $\rho_0$ is the spin stiffness. From hereon we set $c = 1$. The unit vector constraint can be eliminated explicitly by writing $\Vec{n}_0 = (\Vec{\pi}_0,\sigma_0) = (\Vec{\pi}_0,\sqrt{1-\Vec{\pi}_0^2})$. The Lagrangian in terms of the transverse fluctuations $\Vec{\pi}_0$ is
\begin{equation}
    \mathcal{L} = \frac{1}{2} \rho_0 \left[ (\partial_\mu \Vec{\pi}_0)^2 + \frac{(\Vec{\pi}_0 \cdot \partial_\mu \Vec{\pi}_0)^2}{1 - \Vec{\pi}_0^2} \right]. \label{eq:lagrangian_pi}
\end{equation}
Expanding around the zero temperature state of spontaneous symmetry breaking $\sigma_0 = 1$ and $\pi_0^i = 0$ yields
\begin{equation}
    \mathcal{L} = \frac{1}{2} \rho_0 \left[ (\partial_\mu \Vec{\pi}_0)^2 + (\Vec{\pi}_0 \cdot \partial_\mu \Vec{\pi}_0)^2 + \dots \right], \label{eq:lagrangian_quartic}
\end{equation}
where the ellipsis denotes terms of $\mathcal{O}(\pi_0^6)$. The interactions have the effect of renormalizing the spin stiffness---$\rho_0 \rightarrow \rho_\Lambda$---and the fields---$\Vec{n}_0 \rightarrow \Vec{n}_\Lambda \doteq Z_\Lambda^{1/2} \Vec{n}_0$. To see this, we calculate the self-energy by performing a one-loop decoupling of the quartic term
\begin{equation}
    (\Vec{\pi}_0 \cdot \partial_\mu \Vec{\pi}_0)^2 \longrightarrow \frac{1}{N-1} \langle \Vec{\pi}^2  \rangle_\Lambda (\partial_\mu \Vec{\pi}_0)^2,
\end{equation}
where $\langle \Vec{\pi}^2 \rangle_\Lambda$ are the fluctuations of the $\Vec{\pi}$ fields with momenta in the interval $(\Lambda, \Lambda_0)$. At the scale $\Lambda_1 \ll \Lambda_0$ corresponding to several lattice spacings, the fluctuations reduce the effective length of the $\sigma_0$ component down to $n_0 = \lvert\langle\Vec{n}_0\rangle\rvert < 1$. Therefore, we require that the renormalized fields satisfy $\langle \sigma_{\Lambda_1} \rangle = 1$; the renormalized ground state should have the same form $\sigma_{\Lambda_1} = 1$, $\pi^i_{\Lambda_1} = 0$. Far away from the quantum critical point, the first order perturbative calculation gives
\begin{equation}
    Z_{\Lambda_1} = \frac{1}{n_0^2} \simeq 1 + \langle \Vec{\pi}^2 \rangle_{\Lambda_1}, \label{eq:residue_renorm}
\end{equation}
and hence,
\begin{equation}
    \rho_{\Lambda_1} \simeq \rho_0 \left( 1 - \frac{N-2}{N-1} \langle \Vec{\pi}^2  \rangle_{\Lambda_1} \right). \label{eq:stiffness_renorm}
\end{equation}
In practice, $\rho$ and $n_0$ are calculated numerically, which we do using path integral Monte Carlo (see Fig. \ref{fig:rho_mag} and Appendix \ref{app:MC}). By integrating out the ultraviolet quantum fluctuations, we can obtain the low energy ``coarse-grained'' Lagrangian normalized at $\Lambda_1$
\begin{equation}
    \mathcal{L} = \frac{1}{2}\rho(\partial_\mu \Vec{n})^2 , \qquad\qquad \Vec{n}^2 = 1. \label{eq:lagrangian_low_energy}
\end{equation}

Turning to the case of finite temperature, we note that long-range order is destroyed by thermal fluctuations and no state of spontaneously broken symmetry exists \cite{Mermin1966,Hohenberg1967}. However, the correlation length $\xi \propto \exp[2\pi\rho/(N-2)T]$ remains exponentially large in the low temperature regime $T \ll \rho$ \cite{Chakravarty1989,Chubukov1994,Hasenfratz1990a}. Therefore, on momentum scales $\xi^{-1} \ll \Lambda < \Lambda_1$, we can apply the same analysis as above by expanding the low energy Lagrangian \eqref{eq:lagrangian_low_energy} around a \textit{locally} ordered state. The thermal fluctuations $\langle \Vec{\pi}^2  \rangle_\Lambda$ can be evaluated directly when $\Lambda < T < \Lambda_1$:
\begin{align}
    \langle \Vec{\pi}^2  \rangle_\Lambda &= (N-1)\int_\Lambda^{\Lambda_1} \frac{d^2 \mathbf{q}}{(2 \pi)^2} \frac{1}{\omega_{\mathbf{q}} \rho} \frac{1}{e^{\omega_{\mathbf{q}}/T} - 1} \nonumber \\
    &\simeq \frac{(N-1)T}{2 \pi \rho} \log \frac{T}{\Lambda}. \label{eq:thermal_fluct}
\end{align}
Within the Matsubara imaginary time formalism, this same result can be obtained by noting that the dominant contribution comes from the zero Matsubara frequency in the low temperature $T \ll \rho$ regime. This statement is equivalent to the common wisdom that the low temperature regime of the square lattice Heisenberg antiferromagnet is characterized by classical static (zero Matsubara frequency) thermal fluctuations \cite{Chakravarty1989}. Importantly, we note that the ultraviolet cutoff of the logarithm in \eqref{eq:thermal_fluct} is imposed by the Bose occupation factor in the momentum integral.

The renormalization group (RG) equations governing the flow of $\rho$ and $Z$ from the scale $\Lambda_1$ down to $\Lambda$ then follow directly from Eqs. \eqref{eq:residue_renorm}, \eqref{eq:stiffness_renorm} and \eqref{eq:thermal_fluct} by considering an infinitesimally small variation of $l = \log(T/\Lambda)$:
\begin{subequations} \label{eq:rg}
\begin{align}
    &\frac{d \rho}{d l} = - \frac{(N-2)T}{2 \pi}, \label{eq:rg_rho}\\
    & \frac{d \log Z}{d l} = \frac{(N-1)T}{2 \pi \rho}.  \label{eq:rg_residue}
\end{align}
\end{subequations}
When $N \geq 3$, there is a non-trivial renormalization group flow and integrating the system of equations \eqref{eq:rg} yields precisely Eqs. \eqref{eq:RG_equations}.

\subsection{Two-loop contributions to renormalization}

At the end of Sec. \ref{sec:formalism}, we claim that higher-loop order corrections to the spin stiffness and order parameter could be neglected at scales $\xi^{-1} \ll \Lambda < \Lambda_1$. Here, we consider the well-known RG flow equations for the 2D NLSM \cite{Brezin1976,zinn2002quantum}, which describe the interactions of classical fluctuations in the $(2+1)$D model \cite{Chakravarty1989}. Since all calculations should be consistent to leading order with perturbation theory, we again use the temperature $T$ instead of $\Lambda_1$ as the ultraviolet normalization point of the RG flow. The two-loop equations in terms of the dimensionless temperature $t = T/2\pi\rho$ are \cite{Brezin1976},
\begin{subequations}
\begin{align}
    &\frac{dt}{dl} = (N-2)t^2 + (N-2)t^3 + \mathcal{O}(t^4), \\
    & \frac{d \log Z}{d l} = (N-1)t + \mathcal{O}(t^3).
\end{align}
\end{subequations}
Re-writing these differential equations in terms of $\rho$, we find the exact solution
\begin{subequations}
\begin{align}
    \rho_\Lambda &= -\frac{T}{2\pi}\left(1 + W_{-1}\left[ X e^{X} \left(\frac{T}{\Lambda}\right)^{N-2} \right] \right), \label{eq:app:2_loop_rho}\\
    Z_\Lambda &= \frac{1}{n_0^2}\left(\frac{2\pi\rho + T}{2\pi \rho_\Lambda + T} \right)^{\textstyle \frac{N-1}{N-2}},
\end{align}
\end{subequations}
where $X = -(1 + 2\pi\rho/T)$, $W(x)$ is the inverse function of $W e^W = x$---otherwise known as the Lambert $W$ function or product logarithm---which has two branches for real $x$. The $-1$ branch satisfies $W_{-1}(x e^x) = x$ if $x\leq -1$, which guarantees that the initial condition of the differential equation is satisfied.

In the regime we are interested in---$T \ll \rho$ and $\Lambda \gg \xi^{-1}$---the argument of the $W$ function in \eqref{eq:app:2_loop_rho} is close to zero. Using the asymptotic expansion $W_{-1}(x) = \log(-x) - \log(-\log(-x)) + \mathcal{O}(1)$, we find
\begin{subequations}
\begin{align}
    \frac{\rho_\Lambda}{\rho} &= 1 - \frac{(N-2)T}{2\pi\rho} \log\frac{T}{\Lambda} + \mathcal{O}(T^2/\rho^2) , \\
    Z_\Lambda &= \frac{1}{n_0^2}\left(\frac{\rho}{\rho_\Lambda} \right)^{\textstyle \frac{N-1}{N-2}} + \mathcal{O}(T/\rho).
\end{align}
\end{subequations}
Therefore, neither the quasiparticle residue nor the spin stiffness are modified---to a good degree of accuracy---by two-loop contributions in the low temperature regime.

This result may surprise, given that it is well known that the correlation length $\xi$ is heavily modified by two-loop corrections \cite{Sachdev2011,Chakravarty1989}. However, these differences appear only in the far infrared limit. First, observe that the one-loop spin stiffness $\eqref{eq:RG_rho}$ vanishes at the scale $\Lambda = \lambda^{-1}$, where $\lambda$ is given by $\eqref{eq:lambda_scale}$. However, since $W_{-1}(-1/e) = -1$, the two-loop spin stiffness vanishes at the scale $\Lambda = \Xi$, where 
\begin{align}
    \Xi &= T\left(1 + \frac{2\pi\rho}{T}\right)^{\textstyle \frac{1}{N-2}} \exp\left[ - \frac{2\pi\rho}{(N-2)T} \right]  \\
    &= T \left(\frac{2\pi\rho}{T} \right)^{\textstyle \frac{1}{N-2}} \exp\left[ - \frac{2\pi\rho}{(N-2)T} \right] \left[1 + \mathcal{O}\left(\frac{T}{\rho}\right)\right]. \nonumber
\end{align}
It is interesting to compare this expression to the correlation length $\xi$ \eqref{eq:xi_length}. Most importantly, for the case $N = 3$ both $\xi^{-1}$ and $\Xi$ have a temperature independent pre-exponential factor---to leading order in $T/\rho$. This is a considerable difference compared to the temperature dependence of $\lambda$. Obviously, any perturbative calculation like RG is not valid at and beyond the strong coupling scale. However, it is clear from the above analysis that the behavior of the spin stiffness and order parameter are heavily modified by two-loop contributions when approaching that scale.

Finally, we also acknowledge that there are two-loop corrections to the speed $c$. However, the renormalized speed as reported in Ref. \cite{Chubukov1994} is only modified at $\mathcal{O}(T^2/\rho^2)$ when $\Lambda \gg \xi^{-1}$.

\section{Comparison of transverse \& longitudinal spectral responses \label{app:transverse_comparison}}

In Sec. \ref{sec:DSF}, we argue that infinitely-many quanta are created by the source. Here, we justify that this is not simply an artifact of the power series expansion of the square root $\sqrt{1 - \Vec{\pi}^2}$.

\begin{figure}[!t]
    \centering
    \includegraphics[scale=1]{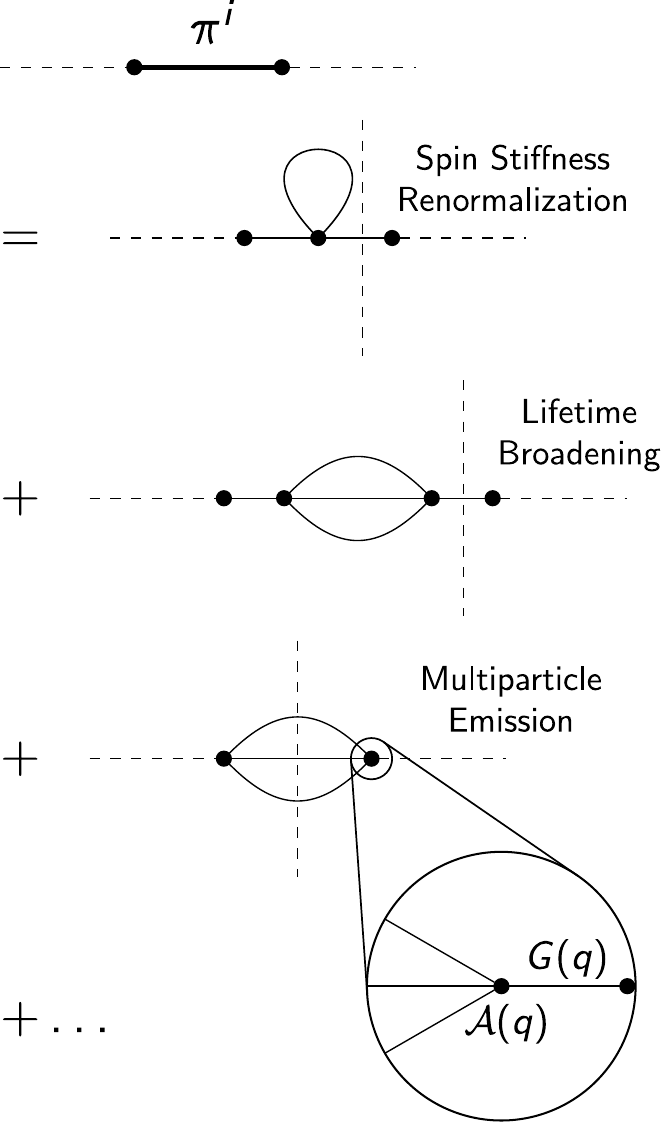}
    \caption{The transverse response of the structure factor also contains infinitely-many multiparticle states. These are obtained from the different ways of cutting diagrams (vertical dashed lines). The thick line represents the exact $\pi^i$ propagator, normal lines represent the bare $\pi^i$ propagator, and horizontal dashed lines represent the source. The inset shows how the propagator $G(q)$ of the virtual magnon contracts with the scattering amplitude $\mathcal{A}(q) = 1/G(q)$ to a single point.}
    \label{fig:multiparticle}
\end{figure}

On the one hand, the DSF must be rotationally invariant. On the other hand, the $n^N = \sqrt{1 - \Vec{\pi}^2}$ longitudinal component contains all even powers of $\Vec{\pi}$, while the transverse components are linear in $\pi^i$. There is no actual contradiction here. To see this, consider Fig. \ref{fig:multiparticle}, where we illustrate the simplest loop corrections to the interaction of the source with a transverse $\pi^i$ component. The first two diagrams correspond to the one- and two-loop contributions to the self energy of the emitted quasiparticle. However, the third diagram shows that the interactions (and self interactions) between the $\pi^i$ components allow the probe to create three real and on-shell particles via an intermediate virtual state. The Feynman rules for the interaction vertex in \eqref{eq:lagrangian_quartic} are given in textbooks (e.g., Ref. \cite{Peskin2018}). In particular, the propagator is $G(q) = i\rho^{-1}/q^2$, and the amplitude for an off-shell particle with three-momentum $q$ to decay into three on-shell particles is $\mathcal{A}(q) = -i\rho q^2 = 1/G(q)$. Therefore, the total quantum amplitude for the intermediate state is $G(q)\mathcal{A}(q) = 1$. Essentially, the virtual particle ``contracts'' the interaction to a single point. 

It is straightforward to see that all higher-order interactions in the expansion of \eqref{eq:lagrangian_pi} will lead to the same amplitude for similar multiparticle emissions. We emphasize that this means that the transverse components also lead to the emission of infinitely-many quasiparticles; whether this infinity is ``odd'' or ``even'' is irrelevant, implying the preservation of $O(N)$ symmetry \cite{OBrien2020}. However, one can see that it is mathematically much simpler to extract the leading contribution to the DSF from the probability of two magnon radiation, as considered in Sec. \ref{sec:DSF}.

\section{Path integral quantum Monte Carlo simulations \label{app:MC}}

In Sec. \ref{sec:equaltime}, we present a subset of our measurements of the equal-time correlation function using path integral quantum Monte Carlo simulations. Here we summarize the details of our simulations---the Monte Carlo update algorithm we implement and how we measure physical observables---and also present a larger selection of data.

\subsection{Heat bath algorithm for \texorpdfstring{$O(3)$}{O(3)} NLSM}

The quantum partition function for the $O(3)$ NLSM in imaginary time is given by the path integral \cite{Chakravarty1989,Manousakis1989},
\begin{subequations}
\begin{align}
    \mathcal{Z} &= \int \mathcal{D}\Vec{n}(\mathbf{x},\tau)\delta(\Vec{n}^2 - 1) e^{-S[\Vec{n}]/\hbar}, \\
    S[\Vec{n}]/\hbar &= \frac{\rho_0}{2\hbar} \int_0^{\hbar/T}d\tau d^2\mathbf{x} \left[\frac{1}{c^2} (\partial_\tau \Vec{n})^2 + (\grad \Vec{n})^2 \right],
\end{align}
\end{subequations}
where $\rho_0$ is the bare, un-renormalized spin stiffness defined at the lattice spacing $b$, $\Vec{n} = (n^{(x)},n^{(y)},n^{(z)})$, and the $\delta$ function in the integration measure enforces the unit vector constraint at every point in space. Discretizing the action over a uniform simple cubic lattice with spacing $b$ yields \cite{Manousakis1989},
\begin{equation}
    S[\Vec{n}]/\hbar = -\frac{1}{g} \sum_{\langle i,j \rangle} \Vec{n}(x_i) \cdot \Vec{n}(x_j), \label{eq:action_discrete}
\end{equation}
where $x_i = (c\tau_i,\mathbf{x}_i)$, $g = \hbar c/\rho_0 b = L_\beta T/\rho_0$, $L_\beta$ is the size (in number of lattice spacings) of the imaginary time dimension, and the summation is over pairs of nearest neighbors. From hereon, we set $\hbar = \rho_0 = b = 1$. In these units, the bare coupling constant $g = c = L_\beta T$.

We performed path integral quantum Monte Carlo simulations of the $O(3)$ model by implementing a heat bath algorithm following Ref. \cite{Manousakis1989}. To summarize:
\begin{enumerate}[label=(\arabic*),wide=0pt,labelindent=\parindent]
    \item Initialize the lattice in a uniformly magnetized grid.
    \item To update a lattice site at position $x_i$, calculate the local action
    \begin{equation}
        \Vec{\omega}(x_i) = \sum_{\langle i,j \rangle} \Vec{n}(x_i) \cdot \Vec{n}(x_j),
    \end{equation}
    so that the probability density for the vector $\Vec{n}(x_i)$ to lie within the solid angle $\Omega$ is
    \begin{equation}
        P(\Omega)d\Omega = C \exp\left(\frac{\lvert\Vec{\omega}\rvert \cos\theta}{g}\right) \sin\theta d\theta d\varphi,
    \end{equation}
    where $C$ is the normalization constant of the distribution, and $\theta$ is measured from the axis directed along $\Vec{\omega}$.
    \item Generate a new configuration for $\Vec{n}(x_i)$ by picking $\theta$ and $\varphi$ from this distribution, convert from local to crystal axis coordinates, and then update.
\end{enumerate}
Defining a ``sweep'' of the lattice to be an update of every lattice site once, we allowed 2500 sweeps for the system to thermalize before starting measurements. We then performed 50,000 sweeps, measuring once every 10 sweeps to minimize correlations between measured configurations; we estimated a correlation time from measurements of the average action per site to be $\approx$ 2 -- 3 sweeps.

\begin{figure*}
    \centering
    \includegraphics[scale=0.7]{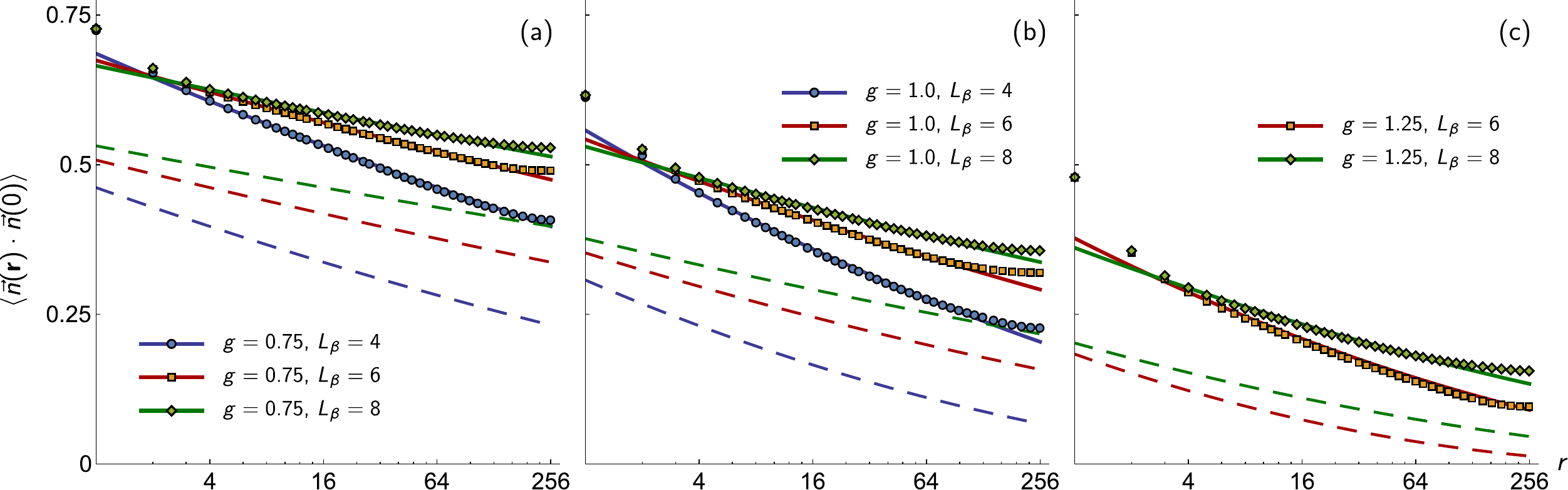}
    \caption{Order parameter equal-time correlation function of the $O(3)$ NLSM measured using Monte Carlo on a $512^2 \times L_\beta$ size lattice with \textbf{(a)} $g=0.75$, \textbf{(b)} $g=1.0$, and \textbf{(c)} $g=1.25$. Zero temperature renormalized parameters and temperature $T = g\rho_0/L_\beta$ are given in Table \ref{tab:zero_temp}. Symbols are Monte Carlo data, solid lines are theory \eqref{eq:equal_time_corr} for $N=3$, and dashed lines are theory replacing $\lambda \rightarrow\xi$ where $\xi$ is the correlation length \eqref{eq:xi_length}. Panel (b) shows the same data as Fig. \ref{fig:correlations}, reproduced here for comparison. All vertical scales are identical. The $L_\beta = 4$ data are omitted from panel (c) since $T/\rho > 1$ is outside the domain of validity of the theory.}
    \label{fig:correlationsA}
\end{figure*}

\begin{figure*}
    \centering
    \includegraphics[scale=0.7]{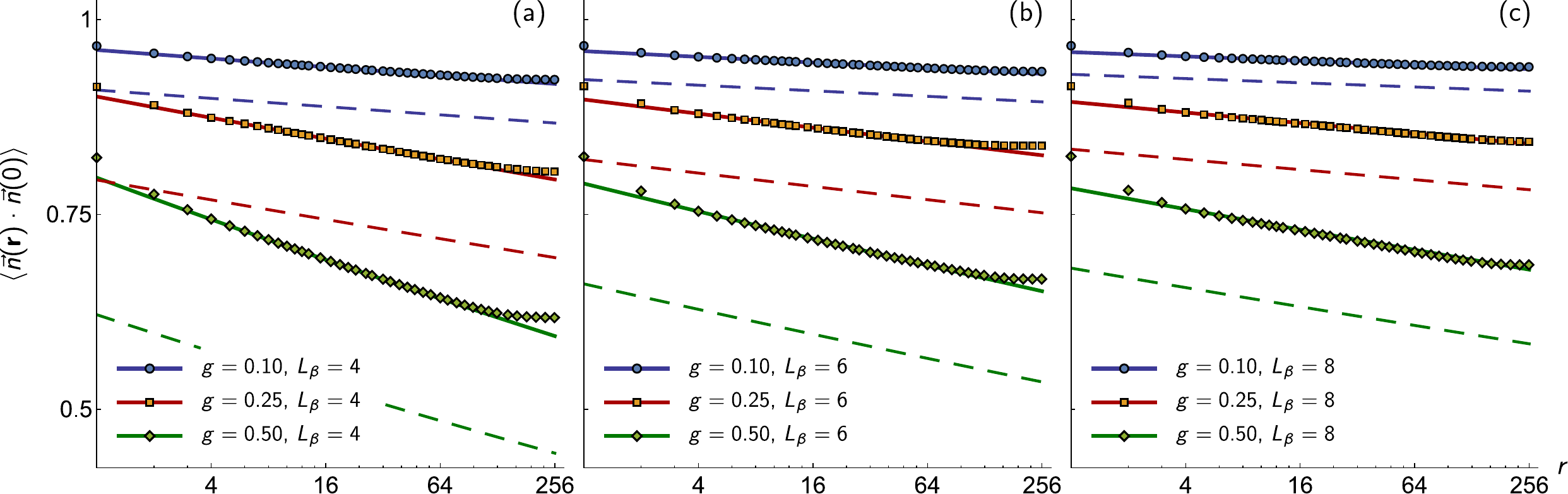}
    \caption{Order parameter equal-time correlation function of the $O(3)$ NLSM measured using Monte Carlo on a $512^2 \times L_\beta$ size lattice with \textbf{(a)} $L_\beta=4$, \textbf{(b)} $L_\beta=6$, and \textbf{(c)} $L_\beta=8$. Zero temperature renormalized parameters and temperature $T = g\rho_0/L_\beta$ are given in Table \ref{tab:zero_temp}. Symbols are Monte Carlo data, solid lines are theory \eqref{eq:equal_time_corr} for $N=3$, and dashed lines are theory replacing $\lambda \rightarrow\xi$ where $\xi$ is the correlation length \eqref{eq:xi_length}. All vertical scales are identical.}
    \label{fig:correlationsB}
\end{figure*}

\renewcommand{\arraystretch}{1.5}
\begin{table*}
    \caption{Selection of zero temperature parameters measured using Monte Carlo on a $L_x  = L_y = L_\beta = 64$ size lattice, the temperature in units of $\rho$, and the length scale $\lambda$ when $L_\beta =$ 4, 6, 8. Statistical fluctuations in measurements are $\mathcal{O}(10^{-5})$.} 
    \begin{ruledtabular}
    \begin{tabular}{ccccccccc}
        &&& \multicolumn{2}{c}{$L_\beta = 4$} & \multicolumn{2}{c}{$L_\beta = 6$} & \multicolumn{2}{c}{$L_\beta = 8$} \\ \cline{4-9}
        $g$ &  $\rho/\rho_0$ & $n_0$ & $T/\rho$ & $\lambda$ & $T/\rho$ & $\lambda$ & $T/\rho$ & $\lambda$ \\ \colrule
        0.10 & 0.958 & 0.975 & 0.026 & $1.5\times10^{105}$ & 0.017 & $4.2\times10^{157}$ & 0.013 & $1.1\times10^{210}$ \\
        0.25 & 0.893 & 0.935 & 0.070 & $3.8\times10^{39}$ & 0.047 & $1.7\times10^{59}$ & 0.035 & $7.2\times10^{78}$ \\
        0.50 & 0.779 & 0.862 & 0.160 & $4.1\times10^{17}$ & 0.107 & $1.9\times10^{26}$ & 0.080 & $8.3\times10^{34}$ \\
        0.75 & 0.651 & 0.779 & 0.288 & $1.2\times10^{10}$ & 0.192 & $9.9\times10^{14}$ & 0.144 & $7.2\times10^{19}$ \\
        1.00 & 0.504 & 0.673 & 0.496 & $1.3\times10^{6}$ & 0.331 & $1.1\times10^{9}$ & 0.248 & $7.9\times10^{11}$ \\
        1.25 & 0.301 & 0.513 & 1.038 & $1.7\times10^{3}$ & 0.692 & $5.3\times10^{4}$ & 0.519 & $1.4\times10^{6}$ \\
    \end{tabular}
    \end{ruledtabular}
    \label{tab:zero_temp}
\end{table*}

\subsection{Measurement methods}

To measure the zero temperature staggered magnetization $n_0$, we used the standard Monte Carlo estimator
\begin{equation}
    n_0^2 = \left\langle \left(\frac{1}{L^3} \sum_i \Vec{n}(x_i) \right)^2 \right\rangle,
\end{equation}
where $L^3$ is the total number of lattice sites (at zero temperature all dimensions are of equal size), and the ensemble average $\langle \,\cdot\, \rangle$ is estimated by an average over measurements.

The equal-time correlation function measurements were obtained using the formula
\begin{align}
    &\langle \Vec{n}(\mathbf{r})\cdot\Vec{n}(0)\rangle = \frac{1}{4 L^2 L_\beta} \nonumber\\
    &\times\bigg\langle  \sum_{i,\mu=x,y} \Vec{n}(x_i) \cdot [\Vec{n}(x_i + r e_\mu) + \Vec{n}(x_i - r e_\mu) ]  \bigg\rangle, 
\end{align}
where $e_\mu$ is a unit vector along the $\mu$ direction, and we averaged over positive and negative displacements along the two equivalent spatial dimensions to improve our measurement statistics; at finite temperature, the imaginary time direction is not equivalent, so is not included in the sum over directions $\mu$. The sum over all lattice sites divided by $L_\beta$ approximates the integral over the imaginary time dimension used to obtain the equal-time correlation function.

To measure the zero temperature renormalized spin stiffness, we adapted the approach described in Ref. \cite{Sandvik2010}, which we summarize here. The spin stiffness measures the response of the system to a twist of the boundary conditions of dimension $\mu$ by a relative angle $\Phi = Q L_\mu$. At zero temperature,
\begin{equation}
    \rho = \frac{1}{L_x L_y} \left. \frac{\partial^2 E(Q)}{\partial Q^2}\right\rvert_{Q=0},
\end{equation}
where $E(Q) = -(g/L_\beta)\log \mathcal{Z}(Q)$ is the ground state energy functional in the presence of the twist and $\mathcal{Z}$ is the quantum partition function. The twisted boundary conditions can be eliminated by transforming to a ``rotating'' frame of reference where the twist instead modifies the local interaction:
\begin{equation}
    S[\Vec{n},Q] = -\frac{1}{g} \sum_{\langle i,j \rangle} \Vec{n}(x_i) \cdot \mathsf{R} \Vec{n}(x_j),
\end{equation}
where $\mathsf{R}$ is a $3\times3$ rotation matrix in spin space. For a rotation about the $\hat{x}$ axis in spin space, along direction $\mu$ in real space, expanding the action to second order in $Q$ leads to a modification of the energy
\begin{equation}
    E(Q) \simeq E(0) - \frac{g}{2 L_\beta} Q^2 \left(\langle S_\mu^{(x)} \rangle + \langle (I_\mu^{(x)})^2 \rangle \right),
\end{equation}
where we have defined
\begin{subequations}
\begin{align}
    S_\mu^{(x)} &= -\frac{1}{g} \sum_{\langle i,j \rangle_\mu} \left[\Vec{n}(x_i) \cdot \Vec{n}(x_j) -  n^{(x)}(x_i) n^{(x)}(x_j) \right],  \\
    I_\mu^{(x)} &= -\frac{1}{g} \sum_{\langle i,j \rangle_\mu} [\Vec{n}(x_i) \times \Vec{n}(x_j)]_x,
\end{align}
\end{subequations}
where summation is over lattice bonds directed in the $\mu$ direction. However, it is necessary to account for the fact that the direction of the twist in spin space is not generally perpendicular to the local magnetization. Therefore, the spin stiffness is obtained by averaging over the other two twist axes, weighted by $3/2$---see the discussion in Sec. \ref{sec:DSF} regarding rotational averaging. At zero temperature, all three Euclidean dimensions are equivalent ($L_x = L_y = L_\beta \doteq L)$, so we also averaged over all bond directions to obtain a more accurate Monte Carlo estimator of the spin stiffness:
\begin{equation}
    \rho = -\frac{g}{3 L^3} \left( \langle S \rangle + \frac{1}{2} \sum_{\mu,a=1}^3 \langle (I_\mu^{(a)})^2 \rangle \right),
\end{equation}
where $\langle S \rangle$ is the average of the action \eqref{eq:action_discrete}.

\subsection{Results \& further analysis}

In Table \ref{tab:zero_temp} we present a subset of measurements of the zero temperature spin stiffness and average staggered magnetization on an $L_x = L_y = L_\beta = 64$ size lattice. These results are practically identical to measurements on a $32^3$ lattice showing that finite size scaling effects are negligible [away from the $O(3)$ quantum critical point $g = g_c \approx 1.46$]. We also present the temperature in units of the renormalized spin stiffness and the length scale $\lambda$. Evidently, as the coupling $g$ is increased, reducing the spin stiffness $\rho$, the relative importance of thermal fluctuations increases.

In Figs. \ref{fig:correlationsA} and \ref{fig:correlationsB} we present measurements of the equal-time correlation function on $L_x = L_y = 512$ and $L_\beta = $ 4, 6, 8 size lattices, for a range of values of the coupling $g$. The solid lines show the theoretical prediction \eqref{eq:equal_time_corr} for $N = 3$ with the zero temperature spin stiffness and magnetization measured on the $64^3$ lattice. We emphasize that the theory has \textit{no adjustable fitting parameters}. Evidently, the agreement between the data and theoretical curves is excellent. As stated in Sec. \ref{sec:equaltime}, disagreement on short length scales ($r \lesssim 2$) is to be expected due to the dominance of ultraviolet quantum fluctuations, and on larger length scales ($r \gtrsim 128$) finite-size effects originating from our choice of periodic boundary conditions become important. Therefore, we see that the Monte Carlo data agrees perfectly with the characteristic length scale $\lambda$, but not at all with the correlation length $\xi$ given by \eqref{eq:xi_length}.

Strictly speaking, our theory is valid in the regime $T \ll \rho$. However, the exponentially-large length scales (both $\lambda$ and $\xi$) mean that it is still possible to study correlations at $r \ll \lambda$ when $T \sim 0.5 \rho$. For example, consider the case $g = 1.25$ and $L_\beta = 6$, shown in Fig. \ref{fig:correlationsA}(c) in red. Here $T/\rho \approx 0.692$ and the theory still agrees quite well with the data. This is because $\lambda \approx $ 53,000 remains more than two orders of magnitude larger than the lattice. In contrast, when $L_\beta = 4$, $T/\rho > 1$ and $\lambda$ is just over twice the (linear) size of the lattice, and as expected the theory \eqref{eq:equal_time_corr} did not agree at all with the data [omitted from Fig. \ref{fig:correlationsA}(c) for clarity] since the temperature is outside the domain of validity.

\bibliography{main}

\providecommand{\noopsort}[1]{}\providecommand{\singleletter}[1]{#1}%
\begin{thebibliography}{24}%
\makeatletter
\providecommand \@ifxundefined [1]{%
 \@ifx{#1\undefined}
}%
\providecommand \@ifnum [1]{%
 \ifnum #1\expandafter \@firstoftwo
 \else \expandafter \@secondoftwo
 \fi
}%
\providecommand \@ifx [1]{%
 \ifx #1\expandafter \@firstoftwo
 \else \expandafter \@secondoftwo
 \fi
}%
\providecommand \natexlab [1]{#1}%
\providecommand \enquote  [1]{``#1''}%
\providecommand \bibnamefont  [1]{#1}%
\providecommand \bibfnamefont [1]{#1}%
\providecommand \citenamefont [1]{#1}%
\providecommand \href@noop [0]{\@secondoftwo}%
\providecommand \href [0]{\begingroup \@sanitize@url \@href}%
\providecommand \@href[1]{\@@startlink{#1}\@@href}%
\providecommand \@@href[1]{\endgroup#1\@@endlink}%
\providecommand \@sanitize@url [0]{\catcode `\\12\catcode `\$12\catcode
  `\&12\catcode `\#12\catcode `\^12\catcode `\_12\catcode `\%12\relax}%
\providecommand \@@startlink[1]{}%
\providecommand \@@endlink[0]{}%
\providecommand \url  [0]{\begingroup\@sanitize@url \@url }%
\providecommand \@url [1]{\endgroup\@href {#1}{\urlprefix }}%
\providecommand \urlprefix  [0]{URL }%
\providecommand \Eprint [0]{\href }%
\providecommand \doibase [0]{https://doi.org/}%
\providecommand \selectlanguage [0]{\@gobble}%
\providecommand \bibinfo  [0]{\@secondoftwo}%
\providecommand \bibfield  [0]{\@secondoftwo}%
\providecommand \translation [1]{[#1]}%
\providecommand \BibitemOpen [0]{}%
\providecommand \bibitemStop [0]{}%
\providecommand \bibitemNoStop [0]{.\EOS\space}%
\providecommand \EOS [0]{\spacefactor3000\relax}%
\providecommand \BibitemShut  [1]{\csname bibitem#1\endcsname}%
\let\auto@bib@innerbib\@empty
\bibitem [{\citenamefont {Banerjee}\ \emph {et~al.}(2017)\citenamefont
  {Banerjee}, \citenamefont {Yan}, \citenamefont {Knolle}, \citenamefont
  {Bridges}, \citenamefont {Stone}, \citenamefont {Lumsden}, \citenamefont
  {Mandrus}, \citenamefont {Tennant}, \citenamefont {Moessner},\ and\
  \citenamefont {Nagler}}]{Banerjee2017}%
  \BibitemOpen
  \bibfield  {author} {\bibinfo {author} {\bibfnamefont {A.}~\bibnamefont
  {Banerjee}}, \bibinfo {author} {\bibfnamefont {J.}~\bibnamefont {Yan}},
  \bibinfo {author} {\bibfnamefont {J.}~\bibnamefont {Knolle}}, \bibinfo
  {author} {\bibfnamefont {C.~A.}\ \bibnamefont {Bridges}}, \bibinfo {author}
  {\bibfnamefont {M.~B.}\ \bibnamefont {Stone}}, \bibinfo {author}
  {\bibfnamefont {M.~D.}\ \bibnamefont {Lumsden}}, \bibinfo {author}
  {\bibfnamefont {D.~G.}\ \bibnamefont {Mandrus}}, \bibinfo {author}
  {\bibfnamefont {D.~A.}\ \bibnamefont {Tennant}}, \bibinfo {author}
  {\bibfnamefont {R.}~\bibnamefont {Moessner}},\ and\ \bibinfo {author}
  {\bibfnamefont {S.~E.}\ \bibnamefont {Nagler}},\ }\bibfield  {title}
  {\bibinfo {title} {{Neutron scattering in the proximate quantum spin liquid
  $\alpha$-RuCl$_3$}},\ }\href {https://doi.org/10.1126/science.aah6015}
  {\bibfield  {journal} {\bibinfo  {journal} {Science}\ }\textbf {\bibinfo
  {volume} {356}},\ \bibinfo {pages} {1055} (\bibinfo {year}
  {2017})}\BibitemShut {NoStop}%
\bibitem [{\citenamefont {Hal{\'{a}}sz}\ \emph {et~al.}(2016)\citenamefont
  {Hal{\'{a}}sz}, \citenamefont {Perkins},\ and\ \citenamefont {van~den
  Brink}}]{Halasz2016}%
  \BibitemOpen
  \bibfield  {author} {\bibinfo {author} {\bibfnamefont {G.~B.}\ \bibnamefont
  {Hal{\'{a}}sz}}, \bibinfo {author} {\bibfnamefont {N.~B.}\ \bibnamefont
  {Perkins}},\ and\ \bibinfo {author} {\bibfnamefont {J.}~\bibnamefont {van~den
  Brink}},\ }\bibfield  {title} {\bibinfo {title} {{Resonant Inelastic X-Ray
  Scattering Response of the Kitaev Honeycomb Model}},\ }\href
  {https://doi.org/10.1103/PhysRevLett.117.127203} {\bibfield  {journal}
  {\bibinfo  {journal} {Physical Review Letters}\ }\textbf {\bibinfo {volume}
  {117}},\ \bibinfo {pages} {127203} (\bibinfo {year} {2016})}\BibitemShut
  {NoStop}%
\bibitem [{\citenamefont {Kastner}\ \emph {et~al.}(1998)\citenamefont
  {Kastner}, \citenamefont {Birgeneau}, \citenamefont {Shirane},\ and\
  \citenamefont {Endoh}}]{Kastner1998}%
  \BibitemOpen
  \bibfield  {author} {\bibinfo {author} {\bibfnamefont {M.~A.}\ \bibnamefont
  {Kastner}}, \bibinfo {author} {\bibfnamefont {R.~J.}\ \bibnamefont
  {Birgeneau}}, \bibinfo {author} {\bibfnamefont {G.}~\bibnamefont {Shirane}},\
  and\ \bibinfo {author} {\bibfnamefont {Y.}~\bibnamefont {Endoh}},\ }\bibfield
   {title} {\bibinfo {title} {{Magnetic, transport, and optical properties of
  monolayer copper oxides}},\ }\href
  {https://doi.org/10.1103/RevModPhys.70.897} {\bibfield  {journal} {\bibinfo
  {journal} {Reviews of Modern Physics}\ }\textbf {\bibinfo {volume} {70}},\
  \bibinfo {pages} {897} (\bibinfo {year} {1998})}\BibitemShut {NoStop}%
\bibitem [{\citenamefont {Mermin}\ and\ \citenamefont
  {Wagner}(1966)}]{Mermin1966}%
  \BibitemOpen
  \bibfield  {author} {\bibinfo {author} {\bibfnamefont {N.~D.}\ \bibnamefont
  {Mermin}}\ and\ \bibinfo {author} {\bibfnamefont {H.}~\bibnamefont
  {Wagner}},\ }\bibfield  {title} {\bibinfo {title} {{Absence of Ferromagnetism
  or Antiferromagnetism in One- or Two-Dimensional Isotropic Heisenberg
  Models}},\ }\href {https://doi.org/10.1103/PhysRevLett.17.1133} {\bibfield
  {journal} {\bibinfo  {journal} {Physical Review Letters}\ }\textbf {\bibinfo
  {volume} {17}},\ \bibinfo {pages} {1133} (\bibinfo {year}
  {1966})}\BibitemShut {NoStop}%
\bibitem [{\citenamefont {Zinn-Justin}(2002)}]{zinn2002quantum}%
  \BibitemOpen
  \bibfield  {author} {\bibinfo {author} {\bibfnamefont {J.}~\bibnamefont
  {Zinn-Justin}},\ }\href@noop {} {\emph {\bibinfo {title} {{Quantum Field
  Theory and Critical Phenomena}}}}\ (\bibinfo  {publisher} {Clarendon},\
  \bibinfo {address} {Oxford},\ \bibinfo {year} {2002})\BibitemShut {NoStop}%
\bibitem [{\citenamefont {Sachdev}(2011)}]{Sachdev2011}%
  \BibitemOpen
  \bibfield  {author} {\bibinfo {author} {\bibfnamefont {S.}~\bibnamefont
  {Sachdev}},\ }\href@noop {} {\emph {\bibinfo {title} {{Quantum Phase
  Transitions}}}}\ (\bibinfo  {publisher} {Cambridge University Press},\
  \bibinfo {address} {Cambridge, England},\ \bibinfo {year} {2011})\BibitemShut
  {NoStop}%
\bibitem [{\citenamefont {Savary}\ and\ \citenamefont
  {Balents}(2017)}]{Savary2017}%
  \BibitemOpen
  \bibfield  {author} {\bibinfo {author} {\bibfnamefont {L.}~\bibnamefont
  {Savary}}\ and\ \bibinfo {author} {\bibfnamefont {L.}~\bibnamefont
  {Balents}},\ }\bibfield  {title} {\bibinfo {title} {{Quantum spin liquids: a
  review}},\ }\href {https://doi.org/10.1088/0034-4885/80/1/016502} {\bibfield
  {journal} {\bibinfo  {journal} {Reports on Progress in Physics}\ }\textbf
  {\bibinfo {volume} {80}},\ \bibinfo {pages} {016502} (\bibinfo {year}
  {2017})}\BibitemShut {NoStop}%
\bibitem [{\citenamefont {O'Brien}\ and\ \citenamefont
  {Sushkov}(2020{\natexlab{a}})}]{OBrien2020a}%
  \BibitemOpen
  \bibfield  {author} {\bibinfo {author} {\bibfnamefont {M.~C.}\ \bibnamefont
  {O'Brien}}\ and\ \bibinfo {author} {\bibfnamefont {O.~P.}\ \bibnamefont
  {Sushkov}},\ }\bibfield  {title} {\bibinfo {title} {{Colossal quasiparticle
  radiation in the Lifshitz spin liquid phase of a two-dimensional quantum
  antiferromagnet}},\ }\href {https://doi.org/10.1103/PhysRevB.101.184408}
  {\bibfield  {journal} {\bibinfo  {journal} {Physical Review B}\ }\textbf
  {\bibinfo {volume} {101}},\ \bibinfo {pages} {184408} (\bibinfo {year}
  {2020}{\natexlab{a}})}\BibitemShut {NoStop}%
\bibitem [{\citenamefont {Chakravarty}\ \emph {et~al.}(1989)\citenamefont
  {Chakravarty}, \citenamefont {Halperin},\ and\ \citenamefont
  {Nelson}}]{Chakravarty1989}%
  \BibitemOpen
  \bibfield  {author} {\bibinfo {author} {\bibfnamefont {S.}~\bibnamefont
  {Chakravarty}}, \bibinfo {author} {\bibfnamefont {B.~I.}\ \bibnamefont
  {Halperin}},\ and\ \bibinfo {author} {\bibfnamefont {D.~R.}\ \bibnamefont
  {Nelson}},\ }\bibfield  {title} {\bibinfo {title} {{Two-dimensional quantum
  Heisenberg antiferromagnet at low temperatures}},\ }\href
  {https://doi.org/10.1103/PhysRevB.39.2344} {\bibfield  {journal} {\bibinfo
  {journal} {Physical Review B}\ }\textbf {\bibinfo {volume} {39}},\ \bibinfo
  {pages} {2344} (\bibinfo {year} {1989})}\BibitemShut {NoStop}%
\bibitem [{\citenamefont {Ty{\v{c}}}\ and\ \citenamefont
  {Halperin}(1990)}]{Ty1990}%
  \BibitemOpen
  \bibfield  {author} {\bibinfo {author} {\bibfnamefont {S.}~\bibnamefont
  {Ty{\v{c}}}}\ and\ \bibinfo {author} {\bibfnamefont {B.~I.}\ \bibnamefont
  {Halperin}},\ }\bibfield  {title} {\bibinfo {title} {{Damping of spin waves
  in a two-dimensional Heisenberg antiferromagnet at low temperatures}},\
  }\href {https://doi.org/10.1103/PhysRevB.42.2096} {\bibfield  {journal}
  {\bibinfo  {journal} {Physical Review B}\ }\textbf {\bibinfo {volume} {42}},\
  \bibinfo {pages} {2096} (\bibinfo {year} {1990})}\BibitemShut {NoStop}%
\bibitem [{\citenamefont {Chubukov}\ \emph {et~al.}(1994)\citenamefont
  {Chubukov}, \citenamefont {Sachdev},\ and\ \citenamefont
  {Ye}}]{Chubukov1994}%
  \BibitemOpen
  \bibfield  {author} {\bibinfo {author} {\bibfnamefont {A.~V.}\ \bibnamefont
  {Chubukov}}, \bibinfo {author} {\bibfnamefont {S.}~\bibnamefont {Sachdev}},\
  and\ \bibinfo {author} {\bibfnamefont {J.}~\bibnamefont {Ye}},\ }\bibfield
  {title} {\bibinfo {title} {{Theory of two-dimensional quantum Heisenberg
  antiferromagnets with a nearly critical ground state}},\ }\href
  {https://doi.org/10.1103/PhysRevB.49.11919} {\bibfield  {journal} {\bibinfo
  {journal} {Physical Review B}\ }\textbf {\bibinfo {volume} {49}},\ \bibinfo
  {pages} {11919} (\bibinfo {year} {1994})}\BibitemShut {NoStop}%
\bibitem [{\citenamefont {Ty{\v{c}}}\ \emph {et~al.}(1989)\citenamefont
  {Ty{\v{c}}}, \citenamefont {Halperin},\ and\ \citenamefont
  {Chakravarty}}]{Tyc1989}%
  \BibitemOpen
  \bibfield  {author} {\bibinfo {author} {\bibfnamefont {S.}~\bibnamefont
  {Ty{\v{c}}}}, \bibinfo {author} {\bibfnamefont {B.~I.}\ \bibnamefont
  {Halperin}},\ and\ \bibinfo {author} {\bibfnamefont {S.}~\bibnamefont
  {Chakravarty}},\ }\bibfield  {title} {\bibinfo {title} {{Dynamic Properties
  of a Two-Dimensional Heisenberg Antiferromagnet at Low Temperatures}},\
  }\href {https://doi.org/10.1103/PhysRevLett.62.835} {\bibfield  {journal}
  {\bibinfo  {journal} {Physical Review Letters}\ }\textbf {\bibinfo {volume}
  {62}},\ \bibinfo {pages} {835} (\bibinfo {year} {1989})}\BibitemShut
  {NoStop}%
\bibitem [{\citenamefont {O'Brien}\ and\ \citenamefont
  {Sushkov}(2020{\natexlab{b}})}]{OBrien2020}%
  \BibitemOpen
  \bibfield  {author} {\bibinfo {author} {\bibfnamefont {M.~C.}\ \bibnamefont
  {O'Brien}}\ and\ \bibinfo {author} {\bibfnamefont {O.~P.}\ \bibnamefont
  {Sushkov}},\ }\bibfield  {title} {\bibinfo {title} {{Anomalous thermal
  broadening from an infrared catastrophe in two-dimensional quantum
  antiferromagnets}},\ }\href {https://doi.org/10.1103/PhysRevB.101.064431}
  {\bibfield  {journal} {\bibinfo  {journal} {Physical Review B}\ }\textbf
  {\bibinfo {volume} {101}},\ \bibinfo {pages} {064431} (\bibinfo {year}
  {2020}{\natexlab{b}})}\BibitemShut {NoStop}%
\bibitem [{\citenamefont {Bloch}\ and\ \citenamefont
  {Nordsieck}(1937)}]{Bloch1937}%
  \BibitemOpen
  \bibfield  {author} {\bibinfo {author} {\bibfnamefont {F.}~\bibnamefont
  {Bloch}}\ and\ \bibinfo {author} {\bibfnamefont {A.}~\bibnamefont
  {Nordsieck}},\ }\bibfield  {title} {\bibinfo {title} {{Note on the Radiation
  Field of the Electron}},\ }\href {https://doi.org/10.1103/PhysRev.52.54}
  {\bibfield  {journal} {\bibinfo  {journal} {Physical Review}\ }\textbf
  {\bibinfo {volume} {52}},\ \bibinfo {pages} {54} (\bibinfo {year}
  {1937})}\BibitemShut {NoStop}%
\bibitem [{\citenamefont {Ioffe}\ and\ \citenamefont
  {Larkin}(1988)}]{Ioffe1988}%
  \BibitemOpen
  \bibfield  {author} {\bibinfo {author} {\bibfnamefont {L.~B.}\ \bibnamefont
  {Ioffe}}\ and\ \bibinfo {author} {\bibfnamefont {A.~I.}\ \bibnamefont
  {Larkin}},\ }\bibfield  {title} {\bibinfo {title} {{Effective Action of a
  Two-Dimensional Antiferromagnet}},\ }\href
  {https://doi.org/10.1142/S0217979288000160} {\bibfield  {journal} {\bibinfo
  {journal} {International Journal of Modern Physics B}\ }\textbf {\bibinfo
  {volume} {02}},\ \bibinfo {pages} {203} (\bibinfo {year} {1988})}\BibitemShut
  {NoStop}%
\bibitem [{\citenamefont {Hohenberg}(1967)}]{Hohenberg1967}%
  \BibitemOpen
  \bibfield  {author} {\bibinfo {author} {\bibfnamefont {P.~C.}\ \bibnamefont
  {Hohenberg}},\ }\bibfield  {title} {\bibinfo {title} {{Existence of
  Long-Range Order in One and Two Dimensions}},\ }\href
  {https://doi.org/10.1103/PhysRev.158.383} {\bibfield  {journal} {\bibinfo
  {journal} {Physical Review}\ }\textbf {\bibinfo {volume} {158}},\ \bibinfo
  {pages} {383} (\bibinfo {year} {1967})}\BibitemShut {NoStop}%
\bibitem [{\citenamefont {Hasenfratz}\ and\ \citenamefont
  {Niedermayer}(1990)}]{Hasenfratz1990a}%
  \BibitemOpen
  \bibfield  {author} {\bibinfo {author} {\bibfnamefont {P.}~\bibnamefont
  {Hasenfratz}}\ and\ \bibinfo {author} {\bibfnamefont {F.}~\bibnamefont
  {Niedermayer}},\ }\bibfield  {title} {\bibinfo {title} {{The exact mass gap
  of the O($N$) $\sigma$-model for arbitrary $N \geq 3$ in $d = 2$}},\ }\href
  {https://doi.org/10.1016/0370-2693(90)90686-Z} {\bibfield  {journal}
  {\bibinfo  {journal} {Physics Letters B}\ }\textbf {\bibinfo {volume}
  {245}},\ \bibinfo {pages} {529} (\bibinfo {year} {1990})}\BibitemShut
  {NoStop}%
\bibitem [{\citenamefont {Nelson}\ and\ \citenamefont
  {Pelcovits}(1977)}]{Nelson1977}%
  \BibitemOpen
  \bibfield  {author} {\bibinfo {author} {\bibfnamefont {D.~R.}\ \bibnamefont
  {Nelson}}\ and\ \bibinfo {author} {\bibfnamefont {R.~A.}\ \bibnamefont
  {Pelcovits}},\ }\bibfield  {title} {\bibinfo {title} {{Momentum-shell
  recursion relations, anisotropic spins, and liquid crystals in 2 + $\epsilon$
  dimensions}},\ }\href {https://doi.org/10.1103/PhysRevB.16.2191} {\bibfield
  {journal} {\bibinfo  {journal} {Physical Review B}\ }\textbf {\bibinfo
  {volume} {16}},\ \bibinfo {pages} {2191} (\bibinfo {year}
  {1977})}\BibitemShut {NoStop}%
\bibitem [{\citenamefont {Lifshitz}\ and\ \citenamefont
  {Pitaevskii}(1980)}]{Lifshitz1995}%
  \BibitemOpen
  \bibfield  {author} {\bibinfo {author} {\bibfnamefont {E.~M.}\ \bibnamefont
  {Lifshitz}}\ and\ \bibinfo {author} {\bibfnamefont {L.~P.}\ \bibnamefont
  {Pitaevskii}},\ }\href@noop {} {\emph {\bibinfo {title} {{Statistical Physics
  Part 2}}}}\ (\bibinfo  {publisher} {Butterworth-Heinemann},\ \bibinfo
  {address} {Oxford},\ \bibinfo {year} {1980})\BibitemShut {NoStop}%
\bibitem [{\citenamefont {Weinberg}(1965)}]{Weinberg1965}%
  \BibitemOpen
  \bibfield  {author} {\bibinfo {author} {\bibfnamefont {S.}~\bibnamefont
  {Weinberg}},\ }\bibfield  {title} {\bibinfo {title} {{Infrared Photons and
  Gravitons}},\ }\href {https://doi.org/10.1103/PhysRev.140.B516} {\bibfield
  {journal} {\bibinfo  {journal} {Physical Review}\ }\textbf {\bibinfo {volume}
  {140}},\ \bibinfo {pages} {B516} (\bibinfo {year} {1965})}\BibitemShut
  {NoStop}%
\bibitem [{\citenamefont {Br{\'{e}}zin}\ and\ \citenamefont
  {Zinn-Justin}(1976)}]{Brezin1976}%
  \BibitemOpen
  \bibfield  {author} {\bibinfo {author} {\bibfnamefont {E.}~\bibnamefont
  {Br{\'{e}}zin}}\ and\ \bibinfo {author} {\bibfnamefont {J.}~\bibnamefont
  {Zinn-Justin}},\ }\bibfield  {title} {\bibinfo {title} {{Spontaneous
  breakdown of continuous symmetries near two dimensions}},\ }\href
  {https://doi.org/10.1103/PhysRevB.14.3110} {\bibfield  {journal} {\bibinfo
  {journal} {Physical Review B}\ }\textbf {\bibinfo {volume} {14}},\ \bibinfo
  {pages} {3110} (\bibinfo {year} {1976})}\BibitemShut {NoStop}%
\bibitem [{\citenamefont {Peskin}\ and\ \citenamefont
  {Schroeder}(1995)}]{Peskin2018}%
  \BibitemOpen
  \bibfield  {author} {\bibinfo {author} {\bibfnamefont {M.~E.}\ \bibnamefont
  {Peskin}}\ and\ \bibinfo {author} {\bibfnamefont {D.~V.}\ \bibnamefont
  {Schroeder}},\ }\href@noop {} {\emph {\bibinfo {title} {{An Introduction To
  Quantum Field Theory}}}}\ (\bibinfo  {publisher} {Westview},\ \bibinfo
  {address} {Boulder, CO},\ \bibinfo {year} {1995})\BibitemShut {NoStop}%
\bibitem [{\citenamefont {Manousakis}\ and\ \citenamefont
  {Salvador}(1989)}]{Manousakis1989}%
  \BibitemOpen
  \bibfield  {author} {\bibinfo {author} {\bibfnamefont {E.}~\bibnamefont
  {Manousakis}}\ and\ \bibinfo {author} {\bibfnamefont {R.}~\bibnamefont
  {Salvador}},\ }\bibfield  {title} {\bibinfo {title} {{Equivalence between the
  nonlinear $\sigma$ model and the spin-1/2 antiferromagnetic Heisenberg model:
  Spin correlations in La$_2$CuO$_4$}},\ }\href
  {https://doi.org/10.1103/PhysRevB.40.2205} {\bibfield  {journal} {\bibinfo
  {journal} {Physical Review B}\ }\textbf {\bibinfo {volume} {40}},\ \bibinfo
  {pages} {2205} (\bibinfo {year} {1989})}\BibitemShut {NoStop}%
\bibitem [{\citenamefont {Sandvik}(2010)}]{Sandvik2010}%
  \BibitemOpen
  \bibfield  {author} {\bibinfo {author} {\bibfnamefont {A.~W.}\ \bibnamefont
  {Sandvik}},\ }\bibfield  {title} {\bibinfo {title} {Computational studies of
  quantum spin systems},\ }\href {https://doi.org/10.1063/1.3518900} {\bibfield
   {journal} {\bibinfo  {journal} {AIP Conference Proceedings}\ }\textbf
  {\bibinfo {volume} {1297}},\ \bibinfo {pages} {135} (\bibinfo {year}
  {2010})}\BibitemShut {NoStop}%
\end{thebibliography}%

\end{document}